\documentclass[aps,prd,reprint,amsmath,amssymb,graphicx,longbibliography,floatfix,nofootinbib]{revtex4-2}

\usepackage{xcolor,graphicx}
\usepackage{subcaption}
\usepackage{hhline} 

\definecolor{darkred}{rgb}{0.4,0.0,0.0}
\definecolor{darkgreen}{rgb}{0.0,0.3,0.0}
\definecolor{darkblue}{rgb}{0.0,0.0,0.7}
\usepackage[bookmarks,linktocpage,colorlinks,
linkcolor = darkred,
urlcolor  = darkgreen,
citecolor = darkblue]{hyperref}

\def\rar{\rightarrow}
\def\l{\langle}
\def\r{\rangle}

\begin{document}

\title{Symmetry Breaking in an Extended O(2) Model}
\author{Leon Hostetler$^{1,2}$}
\email{leonhost@iu.edu\\Present address: Department of Physics, Indiana University, Bloomington, IN 47405}
\author{Ryo Sakai$^{3}$}
\author{Jin Zhang$^{4}$}
\author{Alexei Bazavov$^{2,1}$}
\author{Yannick Meurice$^{5}$}

\affiliation{$^1$ Department of Physics and Astronomy, Michigan State University, East Lansing, MI 48824 USA}
\affiliation{$^2$ Department of Computational Mathematics, Science and Engineering, Michigan State University, East Lansing, MI 48824 USA}
\affiliation{$^3$ Jij Inc., Bunkyo-ku, Tokyo 113-0031, Japan}
\affiliation{$^4$ Department of Physics and Chongqing Key Laboratory for Strongly Coupled Physics, Chongqing University, Chongqing 401331, People's Republic of China}
\affiliation{$^5$ Department of Physics and Astronomy, The University of Iowa, Iowa City, IA 52242 USA }
\def\lt{\lambda ^t}
\def\note{note}
\def\beq{\begin{equation}}
  \def\enq{\end{equation}}

\date{\today}

\begin{abstract}

Motivated by attempts to quantum simulate lattice models with continuous Abelian symmetries using discrete approximations, we study an extended-O(2) model in two dimensions that differs from the ordinary O(2) model by the addition of an explicit symmetry breaking term $-h_q\cos(q\varphi)$. Its coupling $h_q$ allows to smoothly interpolate between the O(2) model ($h_q=0$) and a $q$-state clock model ($h_q\rar\infty$). In the latter case, a $q$-state clock model can also be defined for noninteger values of $q$. Thus, such a limit can also be considered as an analytic continuation of an ordinary $q$-state clock model to noninteger $q$. In previous work, we established the phase diagram for noninteger $q$ in the infinite coupling limit ($h_q\rar\infty$). We showed that there is a second-order phase transition at low temperature and a crossover at high temperature. In this work, we seek to establish the phase diagram at finite values of the coupling using Monte Carlo and tensor methods. We show that for noninteger $q$, the second-order phase transition at low temperature and crossover at high temperature persist to finite coupling. For integer $q=2,3,4$, we know there is a second-order phase transition at infinite coupling (i.e. the well-known clock models). At finite coupling, we find that the critical exponents for $q=3,4$ vary with the coupling, and for $q=4$ the transition may turn into a BKT transition at small coupling. We comment on the similarities and differences of the phase diagrams with those of quantum simulators of the Abelian-Higgs model based on ladder-shaped arrays of Rydberg atoms.

\end{abstract}
\maketitle

\section{Introduction}
\label{sec_intro}

In a quantum many-body system, the Hilbert space grows exponentially with the system size, and the simulation of such systems in the strongly-coupled regime is an extremely hard problem for classical computers. Even for equilibrium systems at Euclidean time, as we have in lattice quantum chromodynamics (QCD), our conventional methods for finite density and real time suffer from sign problems. Real-time dynamics seem completely out of reach unless we use some new approach such as quantum simulation \cite{feynman82}.

To quantum simulate field theories, we need to start with the simplest low-dimensional Abelian models such as the Abelian-Higgs model (i.e. scalar quantum electrodynamics). It has been shown that the Abelian-Higgs model can be mapped to a Rydberg ladder \cite{Bazavov:2015kka, Zhang:2018ufj, meurice2019, Meurice:2021tmab}. In this proposed approach, the Abelian-Higgs Hamiltonian is mapped to an angular momentum Hamiltonian, which in turn is truncated to a finite number of angular momentum states. These can in principle be simulated on a Rydberg ladder, however, one may want to start with even simpler cases. In the limit of infinite self-coupling and zero gauge coupling, the Abelian-Higgs model reduces to the classical O(2) spin model, so the analog simulation of an O(2)-like model is a first step toward the simulation of the Abelian-Higgs model.

In Wilson's formulation of lattice gauge theory, the gauge fields are defined on the links of the lattice. These continuous gauge degrees of freedom give rise to an infinite-dimensional Hilbert space at each link, and it is not obvious how to represent such objects on a quantum simulator. One solution is to truncate the link Hilbert space in some way such that the model with truncated gauge fields still approximates the full theory. For example, in a theory with U(1) gauge fields, the fields may be approximated with $\mathbb{Z}_q$ discrete degrees of freedom \cite{Notarnicola2015}. To optimize such a $\mathbb{Z}_q$ approximation, it is useful to have a continuous family of models that interpolate among the various $\mathbb{Z}_q$ approximations.

In this article, we study the effect of explicitly broken symmetries in classical O(2) spin systems in (1+1) dimensions by adding a symmetry-breaking term
\begin{equation}
\Delta S(h_q, q) = -h_q\sum_x\cos(q\varphi_x)
\end{equation}
to the action of the ordinary O(2) model. 
(Note that the coefficient was denoted $\gamma$ in our previous work~\cite{hostetler2021}, however, we change the notation here to be consistent with the existing literature \cite{JKKN1977, Nguyen2017, Bramwell1997, Taroni2008, Landau1983, Hu1987, Rastelli2004, Calabrese2002, Rastelli2004_q4, Chlebicki2019, Butt2022}.)
When $h_q=0$, the model reduces to the O(2) model, but when $h_q>0$ with $q$ integer, the O(2) symmetry is reduced to a $\mathbb{Z}_q$ symmetry. By taking $h_q \rar \infty$, we recover the $q$-state clock model, and by varying $q$, we can interpolate between the Ising model and the O(2) model. On the other hand, by fixing $q$ and varying $h_q$, we can interpolate between the O(2) model and the $q$-state clock model. We allow also $q$ to be noninteger. For noninteger $q$ and $h_q>0$, the $\mathbb{Z}_q$ symmetry is broken, but there remains a residual $\mathbb{Z}_2$ symmetry, which is associated with an Ising phase transition. By allowing $q$ to take on real values instead of only integer values, one can interpolate among the different $\mathbb{Z}_q$ models.

\begin{figure*}
        \centering
        \begin{subfigure}[b]{0.475\textwidth}
            \centering
            \includegraphics[width=\textwidth]{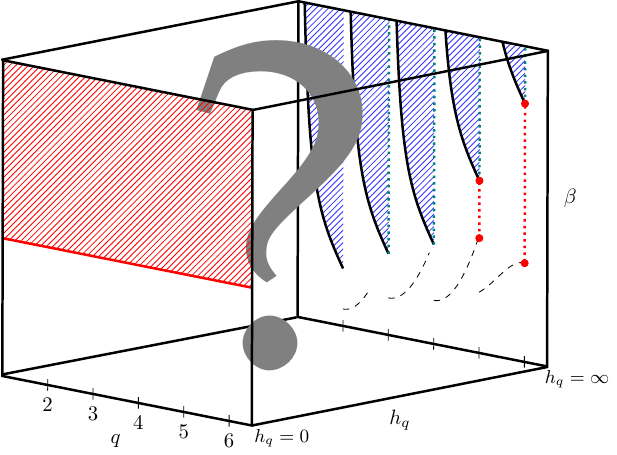}
        \end{subfigure}
        \hfill
        \begin{subfigure}[b]{0.475\textwidth}  
            \centering 
            \includegraphics[width=\textwidth]{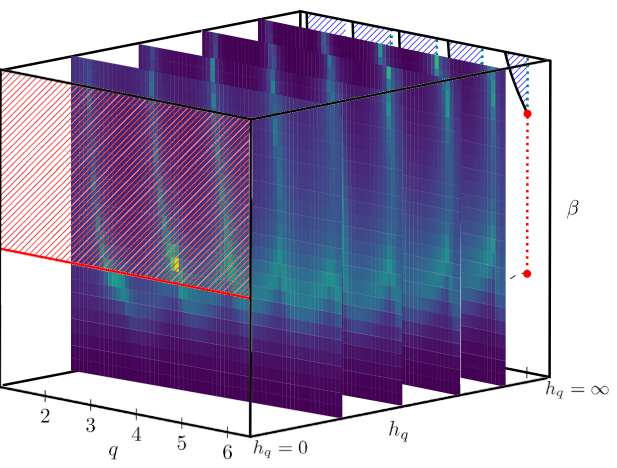}
        \end{subfigure}
        \caption{{\textbf{(Left)} We consider the three-dimensional phase diagram of the Extended-O(2) model for parameter values $h_q,\beta \in \mathbb{R}\geq 0$, and $q\in \mathbb{R}$. In the $h_q=0$ plane, the model reduces to the $XY$ model for all values of $q$. The $XY$ model has a BKT transition near $\beta_c = 1.12$ and a critical phase for $\beta > \beta_c$. In the $h_q=\infty$ plane, the phase diagram is more interesting. For integer $q$, the model reduces to the well-studied clock model, which has a second-order phase transition for $q=2,3,4$ and two BKT transitions for integer $q\geq 5$. For noninteger $q$, we previously established the phase diagram for $h_q\rightarrow\infty$ \cite{hostetler2021}. We found crossovers (dashed lines) at small $\beta$ and second-order phase transitions (solid black lines) of the 2D Ising universality class at large $\beta$. Establishing the phase diagram at finite-$h_q$ is the goal of the present work. \textbf{(Right)} We insert several two-dimensional slices where each slice shows a heatmap, computed using TRG with $L=1024$, of the specific heat at a fixed $h_q$. As described in the main text, such heatmaps can serve as a proxy for the phase diagram. Here they suggest that the sharp edges and corners seen in the $h_q=\infty$ phase diagram become smooth curves at finite $h_q$. However, as we will show later, this does not hold up under scrutiny. A finite-size scaling analysis will show that for noninteger $q$, the phase diagram looks qualitatively the same for all $h_q>0$.}}
        \label{fig:3dphasediagram}
\end{figure*}

In \cite{hostetler2021, hostetler_pos2021}, we studied the $h_q\rar\infty$ limit of this model\footnote{The action Eq.~(\ref{eq_extO2}) differs slightly from the convention of \cite{hostetler2021}, where $\beta$ was treated as a coupling attached to the first term in the action. In the present work, $\beta$ is factored out of the action, so that it plays the role of inverse temperature.}. For integer values of $q$, this is the well-studied clock model. For noninteger $q$, we found a crossover at small $\beta$ and a second-order phase transition of the 2D Ising universality class at larger $\beta$. Thus we extended the phase diagram for the case $h_q\rar\infty$ to noninteger values of $q$. The phase diagram for $h_q=0$ and $h_q=\infty$ is shown in the left panel of Fig.~\ref{fig:3dphasediagram}. In the present work, we study the phase diagram for finite $h_q$. Preliminary results were presented in \cite{hostetler_pos2022}. We performed a scan of the parameter space and computed the specific heat. In many second-order phase transitions the specific heat diverges near the critical point. For such a transition, which is characterized by critical exponent $\alpha>0$, a heatmap of the specific heat can serve as a proxy for the phase diagram. In the right panel of Fig.~\ref{fig:3dphasediagram}, we insert several two-dimensional slices where each slice shows a heatmap of the specific heat at a fixed $h_q$. The heatmaps were computed using tensor renormalization group (TRG) methods developed for the O(2) model \cite{Liu:2013nsa,Yu2014,zhang2021truncation,Meurice:2020pxc} with size $L=1024$. The resulting picture suggests that the sharp edges and corners that one sees at integer values of $q$ in the $h_q=\infty$ phase diagram become smooth curves at finite $h_q$. However, as we will show later, this does not hold up under scrutiny.

\begin{figure}
        \centering
        \begin{subfigure}[b]{0.23\textwidth}
            \centering
            \includegraphics[width=\textwidth]{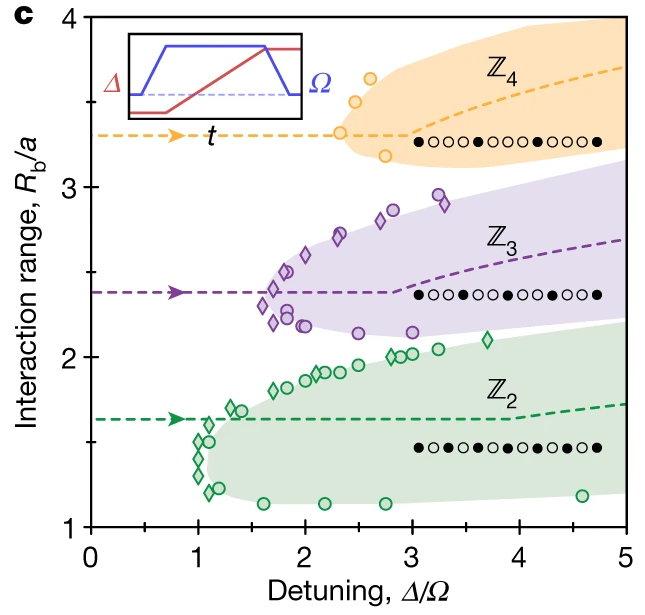}
        \end{subfigure}
        \hfill
        \begin{subfigure}[b]{0.23\textwidth}  
            \centering 
            \includegraphics[width=\textwidth]{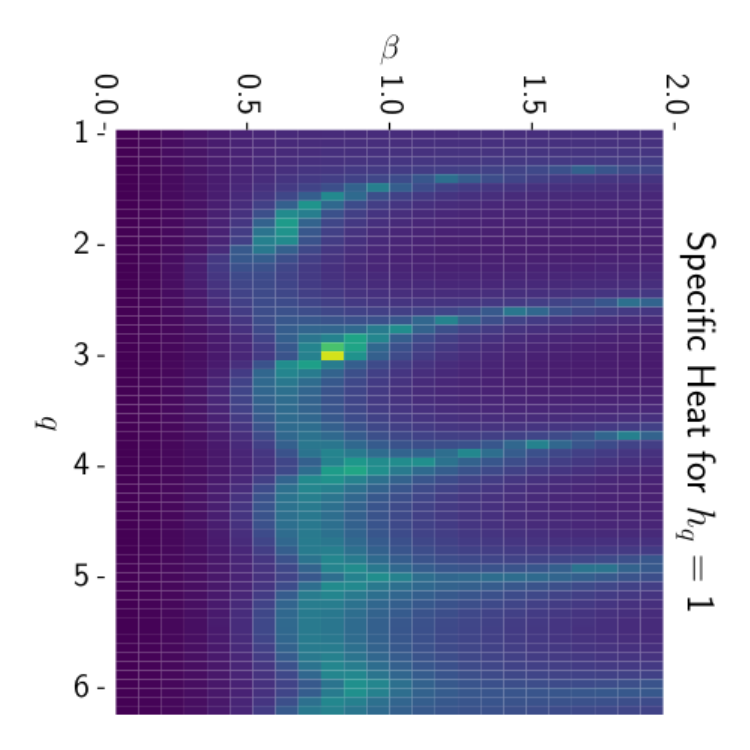}
            \vspace{0.01cm}
        \end{subfigure}
        \caption{{(Left) The phase diagram of a quantum simulator with continuously tunable parameters composed of Rydberg atoms \cite{51qubits, keesling2019}. \textit{Reproduced with permission from Springer Nature.} Note the similarity with a heatmap (right) of the specific heat from the Extended-O(2) model with finite $h_q$.}}
        \label{fig:keesling_comparison}
\end{figure}

The smooth lobes seen in the proxy phase diagram are reminiscent of the phase diagram of a Rydberg atom chain \cite{51qubits, keesling2019} as shown in Fig.~\ref{fig:keesling_comparison}. In the left panel, the phase diagram of a quantum simulator with continuously tunable parameters composed of Rydberg atoms. Note the similarity with a heatmap of the specific heat from the Extended-O(2) model with finite $h_q$. In the Rydberg atom chain, exotic ``floating phases'' are found \cite{Rader2019,PhysRevLett.122.017205,Chepiga&Mila2021Kibble}, 
which gives motivation to search for similar exotic phases in the Extended-O(2) model.

Specific cases of the Extended-O(2) model have been studied before. For $q=2,3$, an early normalization group analysis \cite{JKKN1977} showed that the model at finite $h_q$ should be in the Ising and 3-state Potts universality class respectively. However, Ref.~\cite{Nguyen2017} demonstrated that the critical exponents vary with $h_q$, and concluded that for $q=3$, there is a BKT phase for sufficiently small $h_q$. The $q=4$ with finite $h_q$ case, sometimes referred to as the $XYh_4$ model, has been studied extensively due to its possible relevance to two-dimensional physical systems with 4-fold symmetry such as certain magnetic thin films \cite{Bramwell1997, Taroni2008}. The renormalization group analysis \cite{JKKN1977} showed there to be a second-order phase transition with nonuniversal critical exponents which vary with $h_q$. The BKT phase of the O(2) model was recovered only for $h_q=0$. This is supported by more recent work using a nonperturbative renormalization group approach wherein the model was studied with a continuously varying spatial dimensionality $2 \leq D \leq 3$ \cite{Chlebicki2022}. In Ref.~\cite{Landau1983}, numerical MC studies of the $q=4$ case showed that the critical exponent $\nu$ does vary with $h_q$. This was supported further by Refs.~\cite{Hu1987, Rastelli2004}, and more discussion can be found in Refs.~\cite{Bramwell1997, Calabrese2002}. However, in Ref.~\cite{Rastelli2004_q4}, it was found that for sufficiently small $h_q$, the second-order phase transition for $q=4$ is replaced by BKT transitions. This is supported by Ref.~\cite{Nguyen2017}. However, recent work \cite{Taroni2008, Chlebicki2019} suggests that the apparent BKT phase at finite $h_q$ is essentially a finite-size effect and might not persist to infinite volumes. For $q\geq 5$, the renormalization group analysis \cite{JKKN1977}, showed there should be a BKT phase between a low-temperature ferromagnetic phase and a high-temperature paramagnetic phase. For $q=6$, this is supported by numerical MC studies on triangular lattices \cite{Rastelli2004}. The $q=5$ case was shown to have two transitions for any finite $h_q$ \cite{Butt2022}.

The previous studies of the Extended-O(2) model described above were limited to integer values of $q$, and typically focused only on one or two values. Furthermore, many focused only on large or small values of $h_q$. In the present work, we perform a comprehensive study of the Extended-O(2) model over a large parameter space with $\beta,h_q \in \mathbb{R} \geq 0$ and with $q \in \mathbb{R}$.

In the case $h_q=0$, the Extended-O(2) model reduces to the ordinary O(2) model, which has been studied extensively \cite{Yu2014, Vanderstraeten2019, Gupta1992, Butera1993, Olsson1995, Arisue2009, Komura2012, Jha2020}. In the limit $h_q\rightarrow\infty$, the Extended-O(2) model reduces to the Extended $q$-state clock model, which was studied in \cite{hostetler2021}. With $h_q\rightarrow \infty$ and integer $q$, it reduces to the ordinary $q$-state clock model, which has been studied extensively \cite{Elitzur1979, Rujan1981, Tobochnik1982, Kaski1983, Challa1984, Challa1986, Leroyer1991, Tomita2002, Lapilli2006, Corberi2006, Hwang2009, Baek2009, Baek2010, Borisenko2011, Chatterjee2018, Borisenko2012, Ortiz2012, Chen2017, Chen2018, Surungan2019, Li2020}.

The goal of the present work is to establish the phase diagram of the Extended-O(2) model at finite $h_q$ for both integer and noninteger values of $q$. This is motivated by several things. To quantum simulate Abelian lattice models, it may be necessary to approximate U(1) with, e.g., discrete $\mathbb{Z}_q$ approximations. The Extended-O(2) model with integer $q$ allows to smoothly break the O(2) symmetry of the O(2) model down to the $\mathbb{Z}_q$ symmetry of a clock model. Studying the effect of this symmetry-breaking is crucial to understanding the utility of various $\mathbb{Z}_q$ approximations of $U(1)$. Furthermore, by allowing $q$ to be noninteger, we can explore the effect of breaking the $\mathbb{Z}_q$ symmetry itself, which may be useful for understanding the resilience of a real-life $\mathbb{Z}_q$ approximation to perturbations. Finally, early results from our study of this model suggested a phase diagram reminiscent of one from Rydberg atom chains where exotic floating phases are found. By studying the phase diagram of the Extended-O(2) model at finite $h_q$, we will determine if similar exotic phases can be found in this classical spin system.

This paper is organized as follows. Section~\ref{sec_model} introduces the definition of the Extended-O(2) model and thermodynamic quantities. The Monte Carlo (MC) and TRG methods are introduced in Section~\ref{sec_methods} and a description of the finite-size scaling procedure is given. The results are described in Section~\ref{sec_results}. The phase diagram for integer $q$ is studied using MC and results are given in Section~\ref{sec_integer_q}. For noninteger $q$, the MC approach suffers from very large autocorrelations on larger lattices. Instead, the model is studied using TRG, with results given in Section~\ref{sec_noninteger_q}. The MC method is used to validate the TRG at small volume. In Section~\ref{sec_phase_diagram}, we combine the integer and noninteger $q$ results to present a picture of the phase diagram for the model. In short, we find that for noninteger $q$, there is a crossover at small $\beta$ and an Ising phase transition at larger $\beta$. That is, the picture seen in the limit $h_q\rightarrow\infty$, seems to persist also to finite $h_q$. For integer $q$, the phase diagram is even more interesting. At $h_q\rightarrow\infty$, it is in the clock model universality class with a second-order phase transition for each of $q=2,3,4$ and two BKT transitions for $q\geq 5$. At finite $h_q$, the critical exponents, and hence the universality classes, vary with $h_q$. At small $h_q$, the transition for $q=4$ appears to turn into a BKT transition. We summarize our results in Section~\ref{sec_summary}.

\section{The model}
\label{sec_model}

The classical O(2) spin model is defined on a lattice and has two-component unit vectors on each lattice site. In two-dimensions, the unit vector at a site $x$ can be parameterized by a single angle $\varphi_x\in[0,2\pi)$, and the action takes the form
\begin{align}
\label{eq_O2}
S_{O(2)} = &- \sum_{x,\mu}\cos(\varphi_{x+\hat\mu}-\varphi_{x}) - h \sum_x\cos(\varphi_x-\varphi_h),
\end{align}
where $h = |\vec{h}|$ and $\varphi_h$ are respectively the magnitude and angle of the external magnetic field. We study the ``Extended-O(2) model'' in two dimensions, obtained by adding a symmetry-breaking term to the action of the classical O(2) model
\begin{align}
\label{eq_extO2}
S = &- \sum_{x,\mu}\cos(\varphi_{x+\hat\mu}-\varphi_{x}) - h_q\sum_x\cos(q\varphi_x) \nonumber\\
& \qquad \qquad \qquad - h \sum_x\cos(\varphi_x-\varphi_h).
\end{align}
When $h_q = 0$, this is the ordinary O(2) model. When $h_q>0$, the O(2) symmetry is broken. When $h_q \rightarrow \infty$ and $q$ is integer, the spin angles are forced to take the values $\varphi^{(k)}_{x}=2\pi k/q$ with $k\in\mathbb{Z}$, and hence this becomes the $q$-state clock model. We note that the action defined in Eq.~(\ref{eq_extO2}) is also valid for noninteger $q$.

For noninteger $q$, the $2\pi$-rotational symmetry is broken so we have to carefully define the domain of the angles $\varphi_x$. For example, one could choose $\varphi_x \in [-\pi,\pi)$ or $\varphi_x \in [0,2\pi)$. These different choices lead to very different phase diagrams as we showed for the $h_q\rar\infty$ case in Ref.~\cite{hostetler2021}. We prefer the choice $\varphi_x \in [0,2\pi)$, simply because the phase diagram gives a more consistent periodic picture as $q$ is increased\footnote{See the left panel of Figure 26 in Ref.~\cite{hostetler2021} as opposed to the right panel in the same figure.}.

The partition function of this model is
\begin{equation}
\label{eq_Z}
Z= \int_{-\varepsilon}^{2\pi-\varepsilon} \prod_x\frac{d\varphi _x}{2\pi}e^{-\beta S}.
\end{equation}
where $\varepsilon$ is a small shift of the angles required to connect with the clock models in the $h_q\rightarrow\infty$ limit. Details are provided in Appendix~\ref{app:clockmodellimit}.

We define the internal energy as
\begin{equation}
\label{eq_internalenergy}
\langle E\rangle = -\frac{\partial}{\partial\beta}\ln Z = \left\langle S \right\rangle ,
\end{equation}
where $\langle\dots\rangle$ denotes the ensemble average. We define the specific heat as
\begin{equation}
\label{eq_specificheat}
C_V = \frac{-\beta^{2}}{V} 
\frac{\partial \langle E \rangle}{\partial \beta} = \frac{\beta^{2}}{V} (\langle E^{2} \rangle - \langle E \rangle ^{2}),
\end{equation}
where $V=L^2$ is the lattice volume.

We perform MC simulations with zero external magnetic field ($h=0$). 
Since MC simulations are performed in a finite volume, the expectation value of the magnetization vector $\langle\vec{M}\rangle$ vanishes at zero magnetic field. Therefore, we measure a proxy magnetization
\begin{equation}
\label{eq_proxymag}
\langle|\vec{M}|\rangle=
\left\langle\left|
\sum_x\vec{S}_x
\right|\right\rangle.
\end{equation}
where $\vec{S}_x$ is the two-component unit vector sitting at the lattice site $x$. The corresponding magnetic susceptibility is
\begin{equation}
\label{eq_proxyc}
\chi_{|\vec{M}|}=
\frac{1}{V}\left(\langle |\vec{M}|^2\rangle-
\langle|\vec{M}|\rangle^2\right).
\end{equation}
For the proxy order parameter defined by Eq.~(\ref{eq_proxymag}), we have the corresponding Binder cumulant
\begin{equation}
\label{eq_bincum}
U_{|\vec{M}|} = 1 - \frac{\langle |\vec{M}|^4 \rangle}{3\langle |\vec{M}|^2 \rangle^2}.
\end{equation}
In the limit $\beta\rar\infty$, everything is frozen and so $\langle |\vec{M}|^4 \rangle = \langle |\vec{M}|^2 \rangle^2$, and this Binder cumulant goes to the trivial value $2/3$. In the limit $\beta\rar 0$, the magnetization $\vec{M}$ is a 2-dimensional Gaussian distribution centered at zero. The magnitude $|\vec{M}|$ of such a distribution is itself a Rayleigh distribution, which has a fourth moment $\langle |\vec{M}|^4 \rangle = 2\langle |\vec{M}|^2 \rangle^2$. Hence, in the limit $\beta\rar 0$, this Binder cumulant goes to the trivial value $1/3$. The Binder cumulant varies rapidly with $\beta$ and varies with lattice size. However, it has been shown that at a critical value of $\beta$, the Binder cumulant takes a universal value independent of lattice size \cite{binder1981}, and as such, it is a useful quantity for identifying phase transitions and locating their critical points.

We also consider the Binder cumulant
\begin{equation}
\label{eq_rotatedbincum}
U_{\phi} = 1 - \frac{\langle M_\phi^4 \rangle}{3\langle M_\phi^2 \rangle^2},
\end{equation}
of the ``rotated magnetization" $M_\phi = \cos (q\phi)$, where $\phi \equiv \arctan(M_2/M_1)$, and where $M_1$ and $M_2$ are the first and second components of the magnetization vector $\vec{M}$. In systems with two BKT transitions, e.g. the 5-state clock model, the ordinary Binder cumulant Eq.~(\ref{eq_bincum}) works well to locate the small-$\beta$ critical point but not the large-$\beta$ one. On the other hand, the rotated Binder cumulant Eq.~(\ref{eq_rotatedbincum}) can be used to locate the large-$\beta$ critical point but not the small-$\beta$ one \cite{Baek2009, Baek2010, Borisenko2011, Chatterjee2018}.

We also measure the structure factor
\begin{equation}
\label{eq_sf}
F(\vec{p}) = \frac{1}{L^2}\sum_j \sum_k e^{i(\vec{x}_j - \vec{x}_k)\cdot \vec{p}} \langle \vec{S}_j \cdot \vec{S}_k\rangle,
\end{equation}
which is the Fourier transform of the spin-spin correlator $\langle \vec{S}_j \cdot \vec{S}_k\rangle$. For an $L \times L$ lattice, we define momenta $\vec{p} = 2\pi \vec{n}/L$, with $\vec{n} \in \{(0,0),(0,1),(1,0),\ldots, (L/2,L/2)\}$. The position vector $\vec{x}_j$ is the 2-component Cartesian vector corresponding to the $j$th lattice site. For periodic boundary conditions, we can write Eq.~(\ref{eq_sf}) as
\begin{equation}
\label{eq_sf2}
F(\vec{p}) = \frac{1}{L^2} \left\langle \left|\sum_j e^{-i \vec{p}\cdot \vec{x}_j } \vec{S}_j \right|^2 \right\rangle.
\end{equation}
Now there is only a single sum over the lattice---making this an efficient observable from which to obtain the critical exponent $\eta$ associated with the correlation function.

\section{Methods}
\label{sec_methods}

\subsection{Markov Chain Monte Carlo}

In general, an MCMC updating algorithm involves selecting the next state in the Markov chain from a given probability distribution. In the canonical version of the Metropolis algorithm, a candidate is selected uniformly and followed by an accept/reject step based on the probability of that state occurring in the distribution. For many distributions, the Metropolis algorithm is inefficient due to very low acceptance rate. In contrast, the heatbath algorithm achieves 100\% acceptance by selecting the next state in the Markov chain uniformly from the inverse of the cumulative distribution function (CDF). However, in practice, it is often difficult or impractical to implement a heatbath algorithm because inversion of the CDF may be costly.

At finite $h_q$, the spin angles of the Extended-O(2) model are continuous, and we do not have an efficient heatbath updating algorithm. With the Metropolis algorithm, we found that the acceptance rate dropped as low as $\sim 1\%$. Correspondingly, this lead to large autocorrelations and a critical-like slowing down. To mitigate this, we implemented a biased Metropolis heatbath algorithm \cite{bazavov2005,bazavov2005_2,bazavov2010}, wherein we compute a discretized CDF and use a corresponding biased accept/reject step. This algorithm satisfies the same detailed balance conditions as the unbiased Metropolis algorithm. By decreasing the discretization step size, the acceptance rate is improved but at the cost of increasing the size of the lookup table that is needed. In the limit of infinitesimal discretization step size, one reaches the heatbath acceptance probability of 100\%. We fine-tuned the discretization step size to optimize efficiency which lead to an acceptance rate around $~90\%$.

For the MCMC results, we started with hot lattices (i.e. randomly oriented spins) followed by $2^{15}$ equilibrating sweeps. With our biased Metropolis algorithm, equilibration is much faster than with the unbiased Metropolis algorithm. Measurement sweeps were separated by $2^6-2^{10}$ discarded sweeps. For some lattice sizes and parameter ranges there remained residual autocorrelations which were mitigated by blocking the data with blocksize much larger than the integrated autocorrelation time before calculating means and variances.

For noninteger $q$, the explicitly broken symmetry made it very difficult for MCMC to sample the whole configuration space at low temperature. For lattices larger than $32 \times 32$, the statistics needed become infeasibly large---making it impossible to do the finite-size scaling with MCMC. So for noninteger $q$, the large-volume results were obtained using a tensor renormalization group method.

\subsection{Tensor Renormalization Group}

Using the character expansion
\begin{align}
  e^{\beta \cos(\theta_{1} - \theta_{2})} = \sum_{m=-\infty}^{\infty} I_{m}\left( \beta \right) e^{im(\theta_{1} - \theta_{2})},
\end{align}
where $I$ is the modified Bessel function of the first kind, the partition function $Z$ can be expressed as
\begin{align}
  Z = \sum_{{\{x,t\}}} \prod_{n} A_{x_{n} t_{n} x_{n-\hat{1}} t_{n-\hat{2}}}
\end{align}
with
\begin{align}
\label{eq:Aijkl}
  A_{ijkl} = & \sqrt{I_{i}\left( \beta \right) I_{j}\left( \beta \right) I_{k}\left( \beta \right) I_{l}\left( \beta \right)} \nonumber \\
  & \cdot \int_{0}^{2\pi} \frac{\mathrm{d}\theta}{2\pi} e^{i\theta(i+j-k-l)} e^{\beta h \cos \theta}.
\end{align}
If we do not have the symmetry breaking term $h_q \cos(q\theta)$, there is the ``selection rule'', so that we can analytically evaluate the elements of the tensor.
In the presence of the symmetry breaking term, one can approximate the integral numerically using Gaussian quadrature, which is formulated using orthogonal polynomials. For unbounded integrals, one can use Hermite polynomials as was done in Refs.~\cite{Kadoh:2018tis,Kadoh:2018hqq,Kadoh:2019ube} for non-compact scalar theories.
In our case the integral is bounded, so Legendre polynomials are suitable for the Gaussian quadrature. Legendre polynomials are similarly used in Ref. ~\cite{Kuramashi:2019cgs} for a U(1) model. Then the element of the tensor~\eqref{eq:Aijkl} is evaluated using the quadrature rule
\begin{align}
\int_{0}^{2\pi} \mathrm{d}y f(y)
\approx \pi \sum_{i=1}^{K} w_{i} f\left( \pi x_{i} + \pi \right),
\end{align}
where $x_{i}$ and $w_{i}$ are the $i$-th root of the Legendre polynomial and corresponding weight, respectively, and where $K$ determines the accuracy of this approximation. Note that the elements of $A$ can be complex.

There is another (rather simpler) way to discretize the angles without using the character expansion.
By simply replacing the integral by the summations, we obtain
\begin{align}
  Z \approx \left( \prod_{n} \sum_{\alpha_{n}=1}^{K} \right) \prod_{n} \frac{w_{\alpha_{n}}}{2} \prod_{\nu=1}^{2} h\left( \pi x_{\alpha_{n}} + \pi, \pi x_{\alpha_{n+\hat{\nu}}} + \pi \right).
\end{align}
Now the continuous angle is replaced by the root of the Legendre polynomial that are labeled by discrete $\alpha$ at each site, so that one may be able to consider that $\alpha$s are new discrete d.o.f. of this system.
Then the local Boltzmann factor $h$ can be regarded as a $K \times K$ matrix now and can be decomposed by singular values numerically:
\begin{align}
  \label{eq:svd_h}
  h_{ij} = \sum_{k=1}^{K} U_{ik} \sigma_{k} V^{\dagger}_{kj},
\end{align}
where $\{\sigma\}$ is the singular values and $U$ and $V$ are unitary matrices.
Using them, we can approximately make a tensor network representation of $Z$:
\begin{align}
  Z \simeq \left( \prod_{n} \sum_{\alpha_{n}=1}^{K} \right) \prod_{n} T_{x_{n} t_{n} x_{n-\hat{1}} t_{n-\hat{2}}},
\end{align}
where the tensor is defined by
\begin{align}
  T_{ijkl} = \sum_{\alpha=1}^{K} \frac{w_{\alpha}}{2} \sqrt{\sigma_{i} \sigma_{j} \sigma_{k} \sigma_{l}} U_{\alpha i} U_{\alpha j} V^{\dagger}_{k \alpha} V^{\dagger}_{l \alpha}.
\end{align}
Note that in this construction the elements of $T$ are real. In our work we employ this latter discretization method.

To balance between accuracy and efficiency, one may take a large $K$ and initially truncate the bond dimension of the tensor.
For the current work, $K$ is set to 1024, and the accuracy is examined through a comparison with MC for some parameters (see Appendix~\ref{app:mcvstrg} for several examples).

In this paper, magnetizations and susceptibilities are shown. To calculate them, we employed the method that was proposed by Morita and Kawashima~\cite{Morita2019} and that is based on the higher order TRG (HOTRG). Thus, these quantities are calculated by HOTRG while other quantities like specific heats are calculated by normal (Levin--Nave) TRG.
Note that the significance of the term ``bond dimension'' differs between the TRG and the HOTRG, since the HOTRG gives a better accuracy compared to the normal TRG with a fixed bond dimension. However, also note that we check the convergence and take sufficiently large bond dimensions regardless of which methodology we use.

\subsection{Finite size scaling}

To obtain critical exponents for second-order transitions, we consider the following leading finite size scaling ans{\"a}tze \cite{weigeljanke2010,droz,bartelt}
\begin{align}
\label{eq:ansatz_deru}
\left.\frac{d U_{M}}{d\beta}\right|_{max} &= U_0 + U_1L^{1/\nu} \\
\label{eq:ansatz_cv}
\left.C_V \right|_{max} &= C_0 + C_1 L^{\alpha/\nu} \\
\label{eq:ansatz_inflM}
\left.\langle M \rangle \right|_{infl} &= M_0 + M_1L^{-\beta/\nu} \\
\label{eq:ansatz_magsus}
\left.\chi_{M} \right|_{max} &= \chi_0 + \chi_1 L^{\gamma/\nu} \\
\label{eq:ansatz_sf}
\left.F(\vec{p})\right|_{max} &= F_0 + F_1 L^{2-\eta},
\end{align}
where $\left.\right|_{max}$ means evaluated at the maximum and $\left.\right|_{infl}$ means evaluated at the inflection point. We use $M\equiv|\vec{M}|$ to denote the magnitude of the proxy magnetization. Then $U_{M}$ is the Binder cumulant with respect to the proxy magnetization, $C_V$ is the specific heat, $\chi_M$ is the susceptibility of the proxy magnetization, and $F(\vec{p})$ is the structure factor. Note, when $\alpha=0$, as it is for transitions of the 2D Ising universality class, the ansatz Eq.~(\ref{eq:ansatz_cv}) is modified to
\begin{equation}
\label{eq:ansatz_cv2}
\left.C_V \right|_{max} = C_0 + C_1 \ln L.
\end{equation}

The maxima were extracted by performing an MCMC scan in $\beta$ to locate the approximate position of the peaks. This was followed by a higher resolution scan centered on the peaks. The integrated autocorrelation time of the energy was estimated from each canonical time series. The canonical runs were then ``stitched" together using multihistogram reweighting~\cite{Ferrenberg:1989ui} to interpolate to $\beta$ values between the canonical MCMC runs. The maxima were extracted from jackknife bins by fitting a cubic polynomial to the peak. The jackknife method was used to obtain appropriate error bars for the final results. This was repeated at different lattice sizes, then fitted to the above finite-size scaling ans{\"a}tze in Python using the \textit{curve\_fit} function in the \textit{scipy.optimize} package. More detail is given in Appendix~\ref{app:extractingmaxima}.

For the exponent $1/\nu$, we fit the derivative of the Binder cumulant to the form given by Eq.~(\ref{eq:ansatz_deru}). Instead of taking a finite difference derivative, we follow \cite{weigeljanke2010} and use the analytical form
\begin{align}
\begin{split}
  \label{eq:cumulant_moment}
  \frac{1}{V} \frac{d U_M}{d\beta} &= \frac{2\l M^4\r[\l M^2\r\l E\r-\l M^2E\r]}{3\l M^2\r^3} \\
   & \qquad \qquad -\frac{\l M^4\r\l E\r-\l M^4 E\r}{3\l M^2\r^2}.
\end{split}
\end{align}
where $E$ is the energy density, $M$ is the proxy magnetization, and $V$ is the number of lattice sites.

For the exponent $-\beta/\nu$, we fit Eq.~(\ref{eq:ansatz_inflM}), where $\left.\langle M \rangle \right|_{infl}$ is the value of the proxy magnetization at its inflection point. To determine the location of the inflection point, we use the maximum of the derivative, which can be computed without finite difference as
\begin{equation}
  \label{eq:derm}
  \frac{d \l M \r}{d\beta} = V \left(\l E\r \l M \r - \l EM \r  \right),
\end{equation}
where $E$ is the energy density and $V$ is the number of lattice sites. As with other maxima, a cubic polynomial is fitted to the peak to extract the maximum $\left. d \l M\r / d\beta \right|_{max}$ and its location $\beta_{pc}$. We then interpolate the proxy magnetization to the point $\beta_{pc}$ using a cubic polynomial to obtain $\left.\langle M \rangle \right|_{infl} = \l M \r(\beta_{pc})$.

The quality of each fit was quantified using the $p$-value of the $\chi^2$ statistic as a goodness-of-fit measure:
\begin{equation}
\label{fitq}
p = 1-P( dof/2, \chi^2/2),
\end{equation}
where $P(a,x) = (1/\Gamma(a))\int_0^x t^{a-1}e^{-t} dt$ is the lower incomplete gamma function, and $\chi^2$ is the ordinary chi-squared value of the fit. The $p$-value gives the likelihood that the observed difference between the curve fit and the data is due to chance. A high $p$-value (e.g. $p>0.05$) suggests that the fitting model is correct. We also record the $\chi^2/dof$ for each fit.

The pseudocritical points $\beta_{pc}$ (i.e. the peak locations) for second-order transitions should approach the infinite volume critical point as
\begin{equation}
\label{eq:ansatz_pc}
\beta_{pc} = \beta_{c} + \frac{a}{L^{1/\nu}} .
\end{equation}

To distinguish between second-order and BKT transitions, one can check whether the specific heat diverges with volume. For a BKT transition, the specific heat does not diverge with volume, and so Eq.~(\ref{eq:ansatz_cv}) or (\ref{eq:ansatz_cv2}) is not suitable for BKT transitions.

There are several well-known scaling inequalities \cite{Essam1963, Rushbrooke1963, Josephson1967, Fisher1967, Sokal1981} which relate the different critical exponents. Under certain assumptions, it can be shown that these inequalities become equalities. See any modern statistical physics textbook such as Ref.~\cite{Pathria2022}. In our case, we will consider three such relations:
\begin{align}
\label{eq_scaling1}
\frac{\gamma}{\nu} &= 2- \eta \\
\label{eq_scaling2}
\frac{\alpha}{\nu} &= \frac{2}{\nu}-d \\
\label{eq_scaling3}
d-\frac{\gamma}{\nu} &= 2\frac{\beta}{\nu},
\end{align}
where $d=2$ in our case.

In a second-order transition, the critical exponent $\nu$ is defined by the scaling of the correlation length $\xi \propto t^{-\nu}$, where $t \equiv (T-T_c)/T_c = \beta_c/\beta-1$. However, for a BKT transition, $\xi$ diverges faster than any power of $t$, and so we cannot define $\nu$ in the conventional way. See for example Ref.~\cite{kosterlitz}, where it is shown that for the $XY$ model, $\xi \propto e^{-bt^{-1/2}}$ or \cite{Tobochnik1982}, where it is stated that $\nu$ should be infinite for $q=5,6$.

One can define the exponent $\tilde{\nu}$ for a BKT transition \cite{borisenko} (we use tilde to distinguish from the ordinary $\nu$) by the scaling
\begin{equation}
\label{eq:bkt_nu}
\xi \propto e^{bt^{-\tilde{\nu}}},
\end{equation}
and use this to label the universality class of the system. For the $XY$ model, $\tilde{\nu}=1/2$. Then the pseudocritical points $\beta_{pc}$ (i.e. the peak locations) for BKT transitions should approach the infinite volume critical point as
\begin{equation}
\label{eq:ansatz_pc_bkt}
\beta_{pc} = \beta_{c} + \frac{A}{(\ln L + B)^{1/\tilde{\nu}}} .
\end{equation}
This is different from second-order scaling. However, as noted in Ref.~\cite{borisenko}, fits to this form may be unstable, and so this scaling relation may not be a good way to extract the critical exponent $\tilde{\nu}$. To investigate the $q=5$ state clock model, Ref.~\cite{borisenko} ended up using the equation with $\tilde{\nu}$ fixed at 1/2
\begin{equation}
\label{eq:ansatz_pc_bkt2}
\beta_{pc} = \beta_{c} + \frac{A}{(\ln L + B)^{2}} .
\end{equation}
This form was used to find $\beta_c$, but Ref.~\cite{borisenko} concluded that it was only reliable for very large volumes.

To extract the $\beta$, $\gamma$, and $\eta$ exponents from BKT transitions, we use the scaling ans{\"a}tze Eq.~(\ref{eq:ansatz_inflM}),~(\ref{eq:ansatz_magsus}), and (\ref{eq:ansatz_sf}). For systems with two BKT transitions, e.g. the 5-state clock model, the proxy magnetization does not always work as an order parameter, and so alternative order parameters may have to be used with Eq.~(\ref{eq:ansatz_inflM}) and (\ref{eq:ansatz_magsus}). For example, in the 5-state clock model, the proxy magnetization is useful as an order parameter in the vicinity of the small-$\beta$ critical point, but not the large-$\beta$ one. Near the large-$\beta$ critical point, one can use a ``rotated magnetization'' \cite{Baek2009, Baek2010, Borisenko2011, Chatterjee2018} as is used in Eq.~(\ref{eq_rotatedbincum}).

\section{Results}
\label{sec_results}

\subsection{Integer \texorpdfstring{$q$}{q}}
\label{sec_integer_q}

For integer values of $q$, the Extended-O(2) model retains a rotational symmetry. When $h_q=0$, this is the full O(2) symmetry. When $h_q$ is finite, this symmetry reduces to a $\mathbb{Z}_q$ symmetry. The model is easier to study at integer $q$ because the autocorrelations remain manageable even at large $\beta$.

To get a good survey of the model, we study $q=$ 2, 3, 4, 5, 6 at $h_q=0.1, 1.0$. At infinite $h_q$, the Extended-O(2) model reduces to the ordinary clock model which for $q=2,3,4$ has a single second-order transition (see Table~\ref{tab:clockmodelcriticalpoints234}), and for $q\geq 5$, has a pair of infinite-order BKT transitions. We start with the assumption that the same kind of transitions persist to finite $h_q$. We show then that this assumption fails to hold at intermediate and small $h_q$.

\renewcommand{\arraystretch}{1.3}
\begin{table}
	\centering
	\setlength{\tabcolsep}{6pt}
	\begin{tabular}{cccccccc} \hline
		$q$ & $\alpha$ & $\beta$ & $\gamma$ & $\delta$ & $\eta$ & $\nu$ & $\beta_c$ \\ \hline
		2 & 0 & 1/8 & 7/4 & 15 & 1/4 & 1 & $\frac12 \ln(1+\sqrt{2})$ \\
		3 & 1/3 & 1/9 & 13/9 & 14 & 4/15 & 5/6 & $\frac23 \ln(1+\sqrt{3})$ \\
		4 & 0 & 1/8 & 7/4 & 15 & 1/4 & 1 & $\ln(1+\sqrt{2})$ \\ \hline
	\end{tabular}
	\caption{Here we show (for reference) the critical exponents for the integer $q$ clock model for $q=2,3,4$. The clock model is the $h_q\rightarrow\infty$ limit of the Extended-O(2) model. For $q=2$, this is the Ising model. For $q=3$, the clock model is in the same universality class as the 3-state Potts model. For $q=4$, it has been shown \cite{suzuki} that the clock model is in the 2D Ising universality class. The final column gives the exact infinite-volume critical point for these three cases.}
	\label{tab:clockmodelcriticalpoints234}
\end{table}

We performed MCMC simulations at $L=$ 16, 32, 64, 96, 128, 160, 192, 224, 256. A biased Metropolis heatbath algorithm was used with an acceptance rate around 90\%. Each run started with $2^{15}$ equilibration sweeps followed by $2^{16}$ measurement sweeps. Each measurement sweep was separated by $2^6$ (or more) discarded sweeps to help mitigate autocorrelation. Later during the analysis stage, the residual integrated autocorrelation time was estimated and the effect was removed by binning the observables. For some of the larger volumes, more runs were performed to increase the statistics.

\renewcommand{\arraystretch}{1.1}
\begin{table}
\centering
\begin{tabular}{ccccccc} \hline
$q$ & $h_q$ & $\nu$ & $\alpha$ & $\beta$ & $\gamma$ & $\eta$ \\ \hline
2 & 0.1 & 1.044(53) & 0.021(18) & 0.319(94) & 1.859(96) & 0.287(13) \\
2 & 1.0 & 1.022(71) & -0.010(16) & N/A & 1.78(12) & 0.251(14) \\
3 & 0.1 & 0.658(24) & 0.354(35) & N/A & 1.291(45) & 0.3651(93) \\
3 & 1.0 & 0.809(26) & 0.311(21) & N/A & 1.411(36) & 0.295(16) \\
4 & 0.1 & 2.49(89) & -0.30(12) & 0.76(87) & 4.3(1.5) & 0.2570(85) \\
4 & 1.0 & 1.20(14) & -0.162(19) & 0.78(61) & 2.09(24) & 0.2532(93) \\
5 & 0.1 & N/A & N/A & N/A & N/A & 0.2716(72) \\
5 & 1.0 & N/A & N/A & N/A & N/A & 0.2653(94) \\
6 & 0.1 & N/A & N/A & N/A & N/A & 0.2880(86) \\
6 & 1.0 & N/A & N/A & N/A & N/A & 0.2654(92) \\ \hline
\end{tabular}
\caption{The first two columns refer to the model parameters, and the remaining columns list the critical exponents obtained for the model. They are obtained via the finite-size scaling ans{\"a}tze Eq.~(\ref{eq:ansatz_deru})--(\ref{eq:ansatz_cv2}). See Table~\ref{tab:criticalexponents} in the appendix for more information. Some exponents could not be extracted reliably from our data. For example, in some cases we were unable to get satisfactory fits between our data and Eq.~(\ref{eq:ansatz_inflM}), and so we were unable to extract $\beta/\nu$. For $q=5,6$, we were able to obtain good estimates only for the exponent $\eta$. This is because $q=5,6$ have infinite order BKT transitions for which $\nu$ is not well defined and our method of extracting the bare exponents listed here depends on an estimate of $\nu$ obtained via Eq.~(\ref{eq:ansatz_deru}). This may also be why we see large error bars for the case $q=4$, $h_q=0.1$, which we argue in the main text shows characteristics of a BKT rather than a second-order transition. Note, for $q=5,6$ there are two phase transitions. The results listed here refer to the transition at high temperature.}
\label{tab:bareexponents}
\end{table}

\setlength{\tabcolsep}{0.5em} 
\renewcommand{\arraystretch}{1.1}
\begin{table}
\centering
\begin{tabular}{cccc} \hline
$q$ & $h_q$ & $\nu$ & $\beta_c$ \\ \hline
2 & 0.1 & 1.059(50) & 0.87756(46) \\ 
2 & 1.0 & 0.985(58) & 0.65448(24) \\ 
3 & 0.1 & 0.878(21) & 1.00044(56) \\ 
3 & 1.0 & 0.812(22) & 0.82584(11) \\ 
4 & 0.1 & 2.04(18) & 1.0841(84) \\ 
4 & 1.0 & 1.361(74) & 0.9916(18) \\ 
5 & 0.1 & N/A & 1.136(11) \\ 
5 & 1.0 & N/A & 1.128(12) \\ 
6 & 0.1 & N/A & 1.1251(83) \\ 
6 & 1.0 & N/A & 1.1143(92) \\ \hline
\end{tabular}
\caption{Here we tabulate the jackknife average of the estimates of $\nu$ and $\beta_c$ listed in Tables~\ref{tab:criticalpoints} and \ref{tab:criticalpointsbkt}. These results are obtained from Eq.~(\ref{eq:ansatz_pc}) for $q=2,3,4$ and Eq.~(\ref{eq:ansatz_pc_bkt2}) for $q=5,6$. This provides an alternative estimate of $\nu$ from the one given in Table~\ref{tab:bareexponents} which is obtained from Eq.~(\ref{eq:ansatz_deru}). Note, for $q=5,6$ there are two infinite order BKT phase transitions for which $\nu$ is not well defined. The results listed here refer to the transition at high temperature.}
\label{tab:barenubetac}
\end{table}

Using the FSS fit forms Eq.~(\ref{eq:ansatz_deru})--(\ref{eq:ansatz_cv2}), we extracted the exponents $1/\nu$, $\alpha/\nu$, $-\beta/\nu$, $\gamma/\nu$, and $2-\eta$ for integer $q$ and $h_q=0.1, 1.0$. These are tabulated in Table~\ref{tab:criticalexponents} in Appendix~\ref{app:fss_results_intq}. The critical exponents $\nu$, $\alpha$, $\beta$, $\gamma$, and $\eta$ are then extracted and listed in Table~\ref{tab:bareexponents}. For example, the exponent $\alpha$ is obtained by taking the ratio of measured exponents $\alpha/\nu$ and $1/\nu$. This is done at the level of jackknife bins in order get reliable error bars on the final critical exponents. For $q=5,6$, we do not include results from Eq.~(\ref{eq:ansatz_deru}) since the quantity is unstable and the fits tend to be bad. This is because $q=5,6$ have infinite order BKT transitions for which the critical exponent $\nu$ is not well-defined. As a result, the bare exponents for $q=5,6$, which depend on $1/\nu$ are not listed in Table~\ref{tab:bareexponents}.

For $q=2,3,4$, we fit each observable to Eq.~(\ref{eq:ansatz_pc}) to estimate the critical exponent $\nu$ and critical point $\beta_c$. The results are tabulated in Table~\ref{tab:criticalpoints} in Appendix~\ref{app:fss_results_intq}. Similarly, we use Eq.~(\ref{eq:ansatz_pc_bkt2}), to estimate the BKT critical points for $q=5,6$, and list them in Table~\ref{tab:criticalpointsbkt}. The estimates of $\nu$ and $\beta_c$ are averaged (at the level of jackknife bins) over the different observables to obtain the final estimates listed in Table~\ref{tab:barenubetac}. For example, the first row of Table~\ref{tab:barenubetac} is a jackknife average of the first five rows of Table~\ref{tab:criticalpoints} shown in Appendix~\ref{app:fss_results_intq}. Compare the estimates of the critical exponent $\nu$ in Table~\ref{tab:bareexponents} obtained from Eq.~(\ref{eq:ansatz_deru}) with the estimates in Table~\ref{tab:barenubetac} obtained from Eq.~(\ref{eq:ansatz_pc}). The results are consistent except for the model with $q=3$ and $h_q=0.1$.

\begin{figure}
\centering
\includegraphics{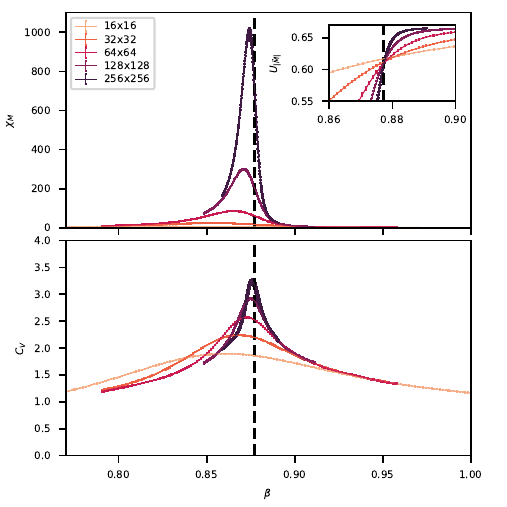}
\caption{Here we look at data for some volumes ($L=16,\ldots,256$) for the model with $q=2$ and $h_q=0.1$. The top panel shows the magnetic susceptibility, the bottom panel shows the specific heat, and the inset within the top panel shows the Binder cumulant. The data is from reweighted Monte Carlo with error bars included. The vertical dashed line is at $\beta_c \approx 0.8773$ where the Binder cumulant curves cross. This figure illustrates the behavior typical of a second-order phase transition---the Binder cumulants cross at the transition point, the magnetic susceptibility diverges, and the specific heat diverges logarithmically.\label{fig:plot_2p0000_g0p1000}}
\end{figure}

For $q=2$ with $h_q = \infty$ (i.e. the Ising model), there is a second-order phase transition at $\beta_c=\ln(1+\sqrt{2})/2 \simeq 0.4407$ with critical exponents $\nu = 1$, $\alpha = 0$, $\gamma = 7/4$, $\beta = 1/8$, and $\eta = 1/4$ (see Table~\ref{tab:clockmodelcriticalpoints234}). With $h_q=1.0$, the thermodynamic quantities show Ising-like divergences near a critical point, which we estimate to be $\beta_c = 0.65448(24)$ in the infinite-volume limit. From finite-size scaling, we find $1/\nu=0.978(68)$, $\alpha/\nu = -0.010(16)$, $\gamma/\nu = 1.745(19)$, and $2-\eta = 1.749(14)$. The bare exponents are $\nu=1.022(71)$, $\alpha=-0.010(16)$, $\gamma=1.78(12)$, and $\eta=0.251(14)$. These are consistent with the critical exponents of the 2D Ising universality class. For this case, we were unable to get satisfactory fits between our data and Eq.~(\ref{eq:ansatz_inflM}), and so we were unable to extract $\beta/\nu$. When we decrease $h_q$ to $0.1$, the thermodynamic quantities show similar Ising-like divergences, with e.g. the specific heat diverging logarithmically (see Fig.~\ref{fig:plot_2p0000_g0p1000}). The critical point is now at $\beta_c = 0.87756(46)$. From finite-size scaling, we find $1/\nu=0.958(49)$, $\alpha/\nu=0.020(17) $, $\gamma/\nu=1.780(16) $, $\beta/\nu=0.306(81) $, and $2-\eta = 1.713(13)$. Our estimates for the bare exponents are $\nu=1.044(53)$, $\alpha=0.021(18)$, $\beta=0.319(94)$, $\gamma=1.859(96)$, and $\eta=0.287(13)$. Our estimates for $\beta$ and $\eta$ are not consistent with the 2D Ising universality class, however, the scaling relations Eq.~(\ref{eq_scaling1}) and (\ref{eq_scaling3}) are violated which may mean that our estimates $\beta/\nu=0.306(81) $, and $2-\eta=1.713(13)$ are not reliable. As $h_q\rightarrow 0$, the critical point moves to larger $\beta$ and connects with the BKT transition near $\beta_c = 1.12$ of the $XY$ model when $h_q=0$.

For $q=3$ with $h_q = \infty$ (i.e. the 3-state Potts model), there is a second-order phase transition at $\beta_c = 2\ln(1+\sqrt{3})/3 \approx 0.6700$, with critical exponents $\nu = 5/6$, $\alpha = 1/3$, $\gamma = 13/9$, $\beta = 1/9$, and $\eta = 4/15$ (see Table~\ref{tab:clockmodelcriticalpoints234}). At $h_q = 1.0$, the thermodynamic functions show second-order divergences near a critical point, which we estimate to be $\beta_c = 0.82584(11)$ in the infinite-volume limit. From finite-size scaling, we find $1/\nu = 1.236(40)$, $\alpha/\nu = 0.385(21)$, $\gamma/\nu = 1.743(16)$, and $2-\eta = 1.705(16)$ which are consistent with the exponents at $h_q=\infty$. The bare exponents are $\nu=0.809(26)$, $\alpha=0.311(21)$, $\gamma=1.411(36)$, and $\eta=0.295(16)$. When we reduce $h_q$ to $0.1$, the critical point moves to $\beta_c = 1.00044(56)$. Assuming a second-order transition, we find from finite-size scaling $1/\nu=1.521(57)$, $\alpha/\nu=0.538(48)$, $\gamma/\nu=1.962(16)$, and $2-\eta=1.6349(93)$, which are different from the 3-state Potts universality class. The bare exponents are $\nu=0.658(24)$, $\alpha=0.354(35)$, $\gamma=1.291(45)$, and $\eta=0.3651(93)$. The scaling relations are violated in this case. As $h_q\rightarrow 0$, the critical point moves to larger $\beta$ and connects with the BKT transition of the $XY$ model when $h_q=0$.

\begin{figure}
\centering
\includegraphics{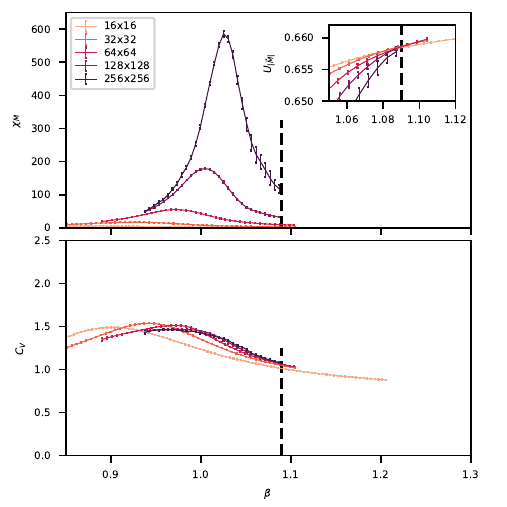}
\caption{Here we look at data for some volumes ($L=16,\ldots,256$) for the model with $q=4$ and $h_q=0.1$. The top panel shows the magnetic susceptibility, the bottom panel shows the specific heat, and the inset within the top panel shows the Binder cumulant. The data is from reweighted Monte Carlo with error bars included. The vertical dashed line is at $\beta_c \approx 1.09$ where the Binder cumulant curves merge. This figure illustrates the behavior typical of a BKT phase transition---the Binder cumulants merge at the transition point, the magnetic susceptibility diverges, and the specific heat plateaus.\label{fig:plot_4p0000_g0p1000}}
\end{figure}

For $q=4$ with $h_q=\infty$, the model is known to be in the 2D Ising universality class with critical point at $\beta_c = \ln(1+\sqrt{2}) \approx 0.8814$ and critical exponents given in Table~\ref{tab:clockmodelcriticalpoints234}. At $h_q=1$, the thermodynamic functions diverge near a critical point which we estimate to be $\beta_c = 0.9916(18)$ in the infinite-volume limit. From finite-size scaling, we find $1/\nu = 0.834(96)$, $\gamma/\nu = 1.742(14)$, $-\beta/\nu=-0.65(51)$, and $2-\eta = 1.7468(93)$. When extracting the exponent $\alpha/\nu$, attempts to fit to the form Eq.~(\ref{eq:ansatz_cv}) failed. However, a fit to the form Eq.~(\ref{eq:ansatz_cv2}) yielded $\alpha/\nu = -0.1349(90)$. The bare exponents are $\nu=1.20(14)$, $\alpha=-0.162(19)$, $\beta=0.78(61)$, $\gamma=2.09(24)$, and $\eta=0.2532(93)$. When we reduce $h_q$ to $0.1$, the critical point moves to $\beta_c = 1.0841(84)$. Assuming a second-order transition, we find from finite-size scaling $1/\nu=0.40(14)$, $\alpha/\nu=-0.120(19)$, $\gamma/\nu=1.728(16)$, $-\beta/\nu=-0.31(31)$, and $2-\eta=1.7430(85)$. The bare exponents are $\nu=2.49(89)$, $\alpha=-0.30(12)$, $\beta=0.76(87)$, $\gamma=4.3(1.5)$, and $\eta=0.2570(85)$. These are not consistent with the 2D Ising universality class. Since $\alpha$ is negative and $\nu$ is finite, the specific heat decreases with increasing volume. This can be seen in Fig.~\ref{fig:plot_4p0000_g0p1000}. Other thermodynamic quantities diverge, and the Binder cumulants merge suggesting a BKT transition instead of a second-order transition. The scaling relation Eq.~(\ref{eq_scaling2}) does not hold in this case, but the other two relations do hold. As $h_q\rightarrow 0$, the critical point moves to slightly larger $\beta$ and connects with the BKT transition of the $XY$ model when $h_q=0$.

\begin{figure}
\centering
\includegraphics{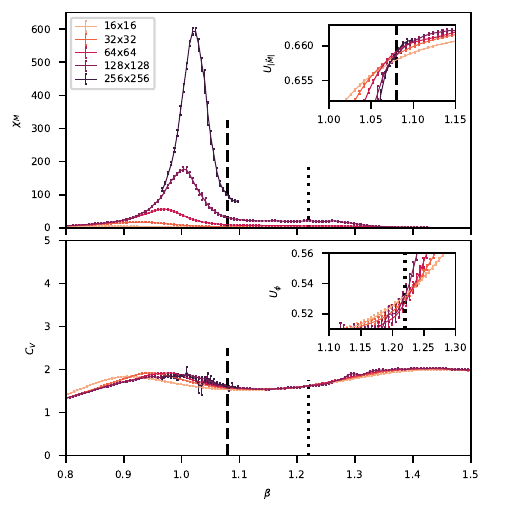}
\caption{Here we look at data for some volumes ($L=16,\ldots,256$) for the model with $q=5$ and $h_q=1.0$. The top panel shows the magnetic susceptibility, and the bottom panel shows the specific heat. The inset within the top panel shows the Binder cumulant of the proxy magnetization. The vertical dashed line is at $\beta_{c1} \approx 1.08$, where the Binder cumulant curves cross. The inset within the bottom panel shows the Binder cumulant of the rotated magnetization defined in Eq.~(\ref{eq_rotatedbincum}). The vertical dotted line is at $\beta_{c2} \approx 1.22$, where the Binder cumulant curves cross. The data is from reweighted Monte Carlo with error bars included. This figure illustrates the behavior typical of the two BKT transitions that occur in the ordinary clock model for $q\geq 5$. The magnetic susceptibility diverges at both transitions, and the specific heat plateaus at both transitions. For $L=256$, we show only results near the first transition point. Already here the error bars are rather large. \label{fig_plot_5p0000_g1p0000}}
\end{figure}

Finally, we consider $q=5,6$. At $h_q=\infty$, these models both show a pair of BKT transitions. At finite $h_q$ there also appear to be two BKT transitions as shown in Fig.~\ref{fig_plot_5p0000_g1p0000}. As before, to extract critical exponents, we look at maxima in the thermodynamic quantities. This works well for the low-$\beta$ transition, but not for the large-$\beta$ transition since many of the thermodynamic quantities do not show a well-defined peak near the second transition. Nevertheless, the transition point can be located by the Binder cumulant crossings. In a BKT transition, the specific heat does not diverge with volume, and one cannot use Eq.~(\ref{eq:ansatz_cv}) to extract a critical exponent $\alpha$. Furthermore, the correlation length diverges faster than any power, and so Eq.~(\ref{eq:ansatz_deru}) cannot be used to extract $\nu$. In fact, in the conventional sense, $\nu$ must be infinite for a BKT transition. Fits to the forms Eq.~(\ref{eq:ansatz_inflM})--(\ref{eq:ansatz_sf}) are performed and exponents recorded in Table~\ref{tab:criticalexponents}. The results are generally consistent with BKT transitions. For $q=5,6$, estimates of the critical points for the small-$\beta$ transition obtained by fitting to Eq.~(\ref{eq:ansatz_pc_bkt2}) are listed in Table~\ref{tab:criticalpointsbkt} in Appendix~\ref{app:fss_results_intq}. The average values are listed in Table~\ref{tab:barenubetac}. However, it is not clear that Eq.~(\ref{eq:ansatz_pc_bkt2}) is a reliable way to estimate the critical point for BKT transitions, and the method of Binder cumulant crossings may be preferred, however this was beyond the scope of this work. For $q=5,6$, the small-$\beta$ transition connects to the BKT transition of the $XY$ model when $h_q = 0$. The large-$\beta$ transition moves to larger $\beta$ as $h_q\rightarrow 0$. Recently, it was estimated that for $q=5$, the large-$\beta$ transition limits to $\beta_{c2} \approx 2.27$ as $h_q \rightarrow 0$ \cite{Butt2022}.

We summarize now our results for integer $q$ and compare them with prior literature. In the infinite-coupling limit $h_q \rightarrow\infty$, this model becomes the ordinary $q$-state clock model. For integer $q$, the clock model has been well-studied and is known to have a second-order transition for $q=2,3,4$ and a pair of BKT transitions for $q \geq 5$. In our previous work \cite{hostetler2021}, we studied this limit of the model, and our results for integer $q$ were consistent with the prior literature. In the present work, we study the model at finite $h_q$. Specific cases appear in the literature, but nobody had yet performed a study of the critical exponents of the model across a large parameter range. For $q=2$, our results are mostly consistent with an Ising phase transition for the entire range of $h_q$ that we studied. This is consistent with an early renormalization group analysis \cite{JKKN1977}. For $q=3$, Ref.~\cite{JKKN1977} predicted the 3-state Potts universality class for all $h_q>0$. Our numerical results agree with this for large $h_q$ but not for small $h_q$, where we find a second-order phase transition with critical exponents different from the 3-state Potts model. Our results for $q=3$ and small $h_q$ are more consistent with a numerical study \cite{Nguyen2017}, which showed that the critical exponents vary with $h_q$. This study also concluded that there is a BKT phase for values of $h_q$ smaller than what we studied here. For $q=4$ with large and intermediate values of the coupling, we find a second-order phase transition with critical exponents that vary with $h_q$. This is consistent with prior literature \cite{JKKN1977, Chlebicki2022, Landau1983, Hu1987, Rastelli2004, Bramwell1997, Calabrese2002}. The literature agrees on what happens at large and intermediate $h_q$, but disagrees on what happens at small $h_q$. The early renormalization group analysis \cite{JKKN1977} showed that the BKT transition is recovered only for $h_q=0$. Numerical studies \cite{Rastelli2004_q4, Nguyen2017} found evidence for BKT transitions at small $h_q$, however, more recent work \cite{Taroni2008, Chlebicki2019} suggests that the apparent BKT phase at finite $h_q$ is essentially a finite-size effect and might not persist to infinite volumes. Our own results for $q=4$ and small $h_q=0.1$ are consistent with BKT, however, our lattices are not large enough to exclude the possibility that these are finite-size effects. Only a few studies by other groups give quantitative results, such as critical points and exponents, which we can compare against. For $q=4$ and $h_q=1.0$, our critical point and critical exponents seem consistent with those in Refs.~\cite{Rastelli2004, Taroni2008}. For $q=4$ and $h_q=0.1$, our results for the critical point and critical exponent $\beta$ agree with those in Ref.~\cite{Rastelli2004_q4}. For $h_q=0.1$ our quantitative results for $\nu$ and $\gamma$ disagree with those in Ref.~\cite{Nguyen2017}, although we agree with the conclusion that the behavior is BKT-like. For integer $q\geq 5$, our results suggest a pair of BKT transitions for all values of $h_q>0$. This is consistent with prior literature including renormalization group analysis \cite{JKKN1977} and numerical studies of specific cases \cite{Rastelli2004, Butt2022}. 

\subsection{Noninteger \texorpdfstring{$q$}{q}}
\label{sec_noninteger_q}

\begin{figure}
\centering
\includegraphics{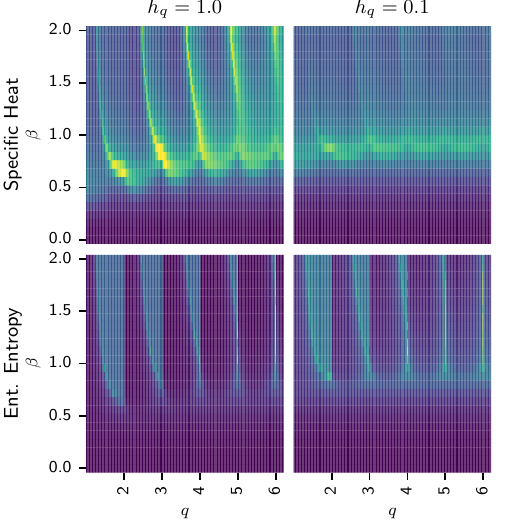}
\caption{Heatmaps from TRG of the specific heat (top row) and entanglement entropy (bottom row) for the Extended-O(2) model with $h_q=1$ (left column) and $h_q=0.1$ (right column). Each point is from a TRG calculation performed with $L=1024$ and bond dimension 40.\label{fig:4panelheatmap}
}
\end{figure}

At noninteger $q$, the explicitly broken $\mathbb{Z}_q$ symmetry causes difficulties for the MCMC approach. In fact, we were unable to get reliable results on lattices larger than $32\times 32$ due to the large autocorrelations that result. This makes finite-size scaling infeasible with the MCMC approach. Instead, we use TRG to investigate the noninteger $q$ regime. We validate the TRG approach by comparison with MCMC on smaller lattices. See Appendix~\ref{app:mcvstrg}.

With TRG, we started with a low precision scan of the parameter space for $q\in[1,6]$, $\beta\in[0,2]$, $h_q=0.1,1.0$, and $L=4,8,\ldots,1024$. For example, we show in Fig.~\ref{fig:4panelheatmap} heatmaps of the specific heat (top row) and entanglement entropy (bottom row) for $h_q=1$ (left column) and $h_q=0.1$ (right column). Each pixel is a data point obtained from TRG performed with $L=1024$ and bond dimension 40. The color in each heatmap ranges from dark to light and this corresponds to a value ranging from 0 to 2.5. The cutoff choice of 2.5 (which truncates some of the specific heat values) was made to increase the contrast in the heatmaps. For $h_q=1$, the heatmap of the specific heat shows smooth lobes even as one crosses integer values of $q$. When $h_q$ is reduced to 0.1, the lines at large $\beta$ fade away leaving a thick horizontal line that connects to the BKT transition at $h_q=0$. Heatmaps of the entanglement entropy (bottom row of Fig.~\ref{fig:4panelheatmap}) show that there remain obvious discontinuities as one crosses integer values of $q$.

The heatmaps of the specific heat show that the heights and positions of the peaks in the specific heat vary smoothly as $q$ is varied. The resulting smooth lobes suggest smooth lines in the phase diagram even as one crosses integer values of $q$. Such a phase diagram is reminiscent of that of Rydberg atom chains as shown in Fig.~\ref{fig:keesling_comparison} and is a motivation for carefully checking for deeper similarities between these two models. On the other hand, heatmaps of the entanglement entropy show that there remain obvious discontinuities as one crosses integer values of $q$. This suggests a phase diagram at finite $h_q$ that is similar to the phase diagram at infinite $h_q$. A rigorous finite-size scaling analysis is needed to establish the true phase diagram at noninteger $q$, and that is what we present in the remainder of this section.

At noninteger values of $q$, the specific heat generally shows two peaks. In the $h_q=\infty$ limit, it was found \cite{hostetler2021} that the small-$\beta$ peak is only a crossover, but the large-$\beta$ peak is associated with a second-order phase transition of the 2D Ising universality class. Here, we investigate the finite $h_q$ regime.

\begin{figure}
\centering
\includegraphics[scale=1]{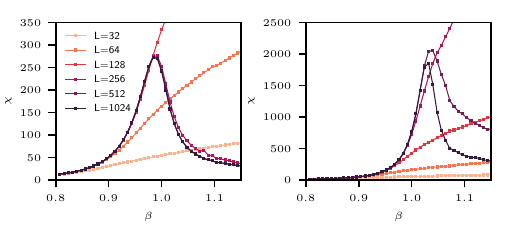}
\caption{Here we look at the behavior of the magnetic susceptibility near the small-$\beta$ peak for noninteger $q$. In this example, $q=4.1$. We obtain the susceptibility from TRG with bond dimension 32 after applying small external magnetic fields $h=10^{-5}$, $\delta h = 10^{-5}$. In the left panel, $h_q = 0.01$ and in the right panel, $h_q = 0.001$. For sufficiently large volumes, the magnetic susceptibility peaks plateau---implying a crossover. For a true phase transition, the peaks would diverge with increasing volume. As the symmetry-breaking parameter $h_q$ is decreased toward zero, one must go to larger volumes to show that the peaks are not diverging.
\label{trgsusccollapse_q4.1}}
\end{figure}

First we consider the small-$\beta$ peak, which we know \cite{hostetler2021} is a crossover at $h_q=\infty$. On the other hand, this peak connects to the BKT transition of the $XY$ model when $h_q=0$. The question is; what happens at finite $h_q$? We find that the magnetic susceptibility plateaus with volume---indicating a crossover---since a phase transition here would result in divergence of the susceptibility with volume (see Fig.~\ref{trgsusccollapse_q4.1}). We conclude that the small-$\beta$ peak corresponds to a crossover even as $h_q\rightarrow 0$ and only becomes a BKT transition exactly at $h_q=0$.

\begin{figure}
\centering
\includegraphics[scale=1]{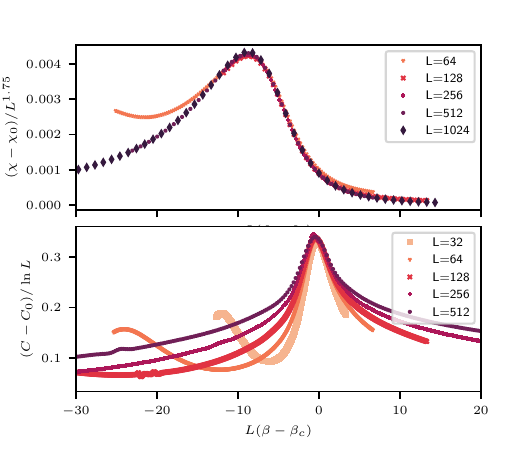}
\caption{In the top panel, the data collapse of the magnetic susceptibility from TRG for the large-$\beta$ peak for the model with $q=3.9$ and $h_q=1$. The estimate $\beta_c=1.196$ was obtained by fitting the peak positions to the finite-size scaling form Eq.~(\ref{eq:ansatz_pc}). To extract the magnetic susceptibility, an external field $h=40/L^{15/8}$ was imposed (with $dh=10^{-5}$ for the numerical differentiation). In the bottom panel, the data collapse of the specific heat for the same model. These results are consistent with there being a second-order phase transition here of the 2D Ising universality class. \label{trg_data_collapse}}
\end{figure}

Next, we consider the large-$\beta$ peak, which corresponds to a second-order transition of the 2D Ising universality class when $h_q=\infty$. As $q$ approaches an integer from below, the second-order transition (i.e. the second peak) occurs at relatively small values of $\beta$. As $q$ is decreased toward the next integer, the critical point moves to large values of $\beta$. In Fig.~\ref{trg_data_collapse}, we show TRG results for $q=3.9$ and $h_q=1$ near the large-$\beta$ peak. In the top panel, we show the data collapse of the magnetic susceptibility from TRG. The estimate $\beta_c=1.196$ was obtained by fitting the peak positions to the finite-size scaling form Eq.~(\ref{eq:ansatz_pc}). To extract the magnetic susceptibility, an external field $h=40/L^{15/8}$ was imposed (with $dh=10^{-5}$ for the numerical differentiation). In the bottom panel, we show the data collapse of the specific heat from TRG for the same model. The specific heat was computed from the second derivative of the free energy after applying smoothing splines to the free energy. These data collapse curves are consistent with a second-order phase transition of the 2D Ising universality class.

In summary, for all noninteger $q$ and all $h_q>0$, we find a crossover at small-$\beta$ and a phase transition of the Ising universality class at large $\beta$. As $q$ approaches an integer from below, the Ising transition line smoothly connects to the critical point of the model with integer $q$. However, for $q$ a little larger than the integer, the Ising transition occurs at much larger $\beta$. 
The apparent similarity with the phase transition of a Rydberg atom chain therefore does not hold up under scrutiny. In fact, the true phase diagram for noninteger $q$ looks qualitatively the same for all $h_q>0$ with sharp discontinuities at integer values of $q$.

\subsection{Phase Diagram}
\label{sec_phase_diagram}

\begin{figure*}
\centering
\includegraphics[scale=0.64]{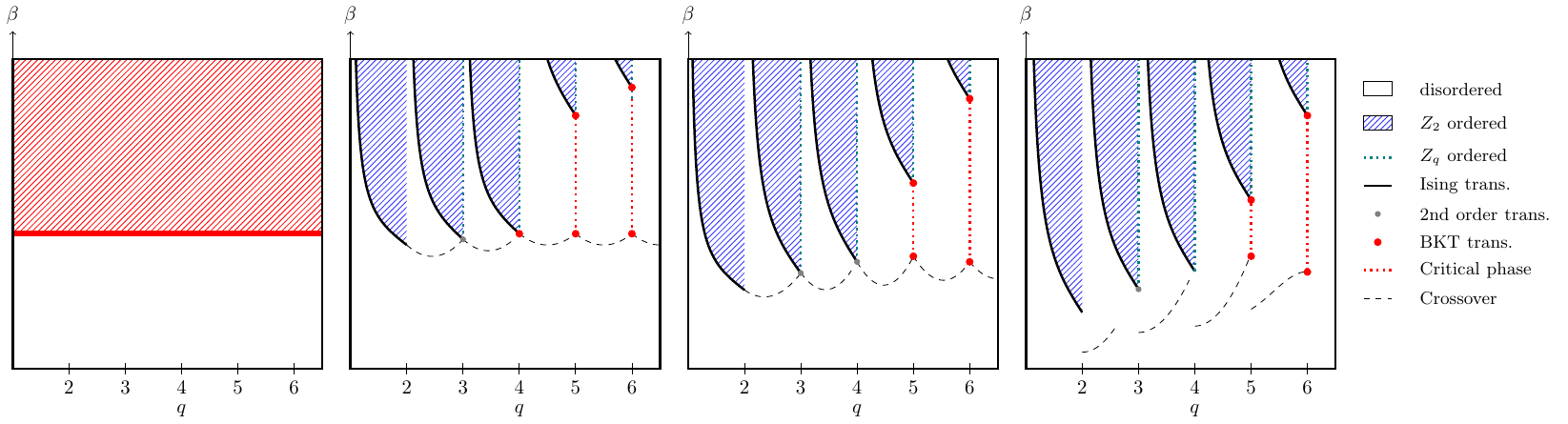}
\caption{Here we present four two-dimensional scenarios of the full three-parameter phase diagram. Each slice is at a different value of the symmetry-breaking parameter $h_q$. In the left-most figure, the phase diagram at $h_q=0$. This is the well-studied $XY$ model at all values of $q$. There is a single BKT transition near $\beta_c=1.12$ and a BKT critical phase at $\beta\geq\beta_c$. The second and third figures give the conjectured phase diagram at small and intermediate $h_q$ respectively. In the right-most figure, we show the phase diagram for the $h_q\rightarrow \infty$ case. In this limit, the model reduces to the $q$-state clock model which has been extended to noninteger $q$ \cite{hostetler2021}. For $h_q>0$ we find for noninteger $q$ a crossover at small $\beta$ and an Ising phase transition at larger $\beta$ with an Ising ordered phase at large $\beta$. As $h_q$ is increased, the only qualitative change for noninteger $q$ is that everything shifts to smaller $\beta$. For $q=2$, our results are consistent with an Ising transition for all $h_q>0$. For $q=3$, our results are consistent with a second-order phase transition with critical exponents that vary with $h_q$. For $q=4$, which is in the 2D Ising universality class in the clock model limit ($h_q\rightarrow\infty$), we find at intermediate $h_q$ a second-order phase transition that may be in a different universality class. At small $h_q$, our results are consistent with a BKT transition for $q=4$. For integer $q\geq 5$, there are two BKT transitions with the small-$\beta$ transitions connecting to the BKT line at $h_q=0$ and the large-$\beta$ transitions limiting to some other points in the $h_q\rightarrow 0$ limit.
\label{pd_allcombined}}
\end{figure*}

Combining our findings for integer and noninteger $q$, we obtain the phase diagram shown in Fig.~\ref{pd_allcombined}. For $h_q = 0$, we have the $XY$ model, which has a BKT phase transition, for all values of $q$. For $h_q\rightarrow\infty$, we have the phase diagram of the $q$-state clock model which has been extended to noninteger values of $q$.

In the limit $h_q\rightarrow\infty$, for $q=2,3,4$, there is a single second-order phase transition of the $q$-state clock model universality class, and for integer $q\geq 5$, there are two BKT phase transitions. When $h_q$ is dialed toward zero, the two BKT transitions for $q\geq 5$ appear to persist to all $h_q>0$. The small-$\beta$ BKT transitions merge with the BKT transition of the $XY$ model at $h_q=0$. The large-$\beta$ BKT transitions, which vanish only for $h_q=0$, converge to different but finite values of $\beta$ in the limit $h_q\rightarrow 0$ \cite{Butt2022}. For $q=3,4$, the critical exponents of the second-order phase transitions vary with $h_q$. We see evidence that the transition for $q=4$ turns into a BKT phase transition at small $h_q>0$. There is some controversy about this point with some authors claiming there is a BKT transition at small $h_q$ \cite{Rastelli2004_q4, Nguyen2017} and others arguing that the BKT transition only occurs for $h_q=0$ \cite{JKKN1977, Taroni2008, Chlebicki2019}. In the present work, our lattices are not large enough to conclusively answer this question.

For noninteger $q$, we find a crossover at small $\beta$, and a second-order phase transition of the 2D Ising universality class at large $\beta$ for all $h_q>0$. This holds for all noninteger $q>1$, so between each consecutive pair of integers $\lfloor q\rfloor$ and $\lceil q\rceil$, there is a smooth second-order transition line that terminates at the integer $\lceil q\rceil$. Thus, the phase diagram is discontinuous at each integer $q$. In this model, the symmetry-breaking term pushes the spins into the angles $\varphi = 2\pi k/q$ where $k=0,1,\ldots, \lfloor q\rfloor$. In some sense the model ``prefers'' these angles with the strength of that preference being given by the coupling strength $h_q$. As $q$ is dialed across an integer value, an additional preferred angle becomes available, so it is not surprising to see discontinuities in the phase diagram at each integer $q$ when $h_q$ is large. It is more surprising that this picture for noninteger $q$ persists to all finite $h_q>0$. That is, contrary to the picture suggested by heatmaps of the specific heat, the discontinuities in the phase diagram persist even to very small $h_q$. More details can be found in Ref.~\cite{Hostetler2023Thesis}.

\section{Summary}
\label{sec_summary}

We studied the effects of adding a symmetry breaking term $-h_q\sum_x \cos(q\varphi_x)$ to the energy function of the ordinary classical O(2) model in two dimensions. This adds two continuously tunable parameters which allows us to explore the effect of symmetry-breaking in O(2)-like models. There is some previous literature on this model, however, it is generally limited to specific integer values of $q$ and specific values of $h_q$. In this work, we have studied a much larger parameter space, and unlike previous studies, we considered also noninteger values of $q$ in order to fully explore the effects of broken symmetries. The result is a rich phase diagram illustrated in Fig.~\ref{pd_allcombined}.

When $q$ is integer, the symmetry-breaking term results in the O(2) symmetry being reduced to a $\mathbb{Z}_q$ symmetry. In the $h_q\rightarrow\infty$ limit, the model reduces to the ordinary $q$-state clock model, which for $q=2,3,4$ has a second-order phase transition and for $q\geq 5$ has two BKT transitions. In the limit $q\rightarrow\infty$, the ordinary O(2) model is recovered. For $q\geq 5$, the pair of BKT transitions seems to persist to finite $h_q$. When $h_q\rightarrow 0$, the small-$\beta$ transition connects to the BKT transition of the O(2) model, whereas the large-$\beta$ transition goes to a different finite value of $\beta$. For $q=3,4$, the phase diagram changes more significantly as the symmetry-breaking parameter $h_q$ is changed. For $q=3$ at small $h_q$ and for $q=4$ at intermediate $h_q$, we find second-order phase transitions with critical exponents different from those in the $h_q\rar\infty$ limit. For $q=4$, our results indicate that the transition turns into a BKT transition at small $h_q$.

For noninteger $q$, early results at finite $h_q$ suggested smooth ``lobes'' of $\mathbb{Z}_q$ ordered phases in the phase diagram reminiscent of those seen in Rydberg atom chains (see Fig.~\ref{fig:keesling_comparison}). However, this initial picture did not hold up under scrutiny. In the $h_q\rightarrow\infty$ limit, we find that for noninteger $q$, there is a crossover at small $\beta$ and an Ising phase transition at larger $\beta$. In the phase diagram, there is a smooth Ising phase transition line that terminates at the next larger integer, such that the phase diagram looks nearly periodic from integer to integer, but with an abrupt discontinuity at each integer. For noninteger $q$, this picture seems to persist to all finite $h_q>0$. In the phase diagram, Fig.~\ref{pd_allcombined}, we see numerous examples of Ising transition lines (noninteger $q$) terminating at BKT transition points (integer $q$) and followed by crossover lines on the other side of the integer. This vividly illustrates the strong influence that symmetries exert in these classes of models and that even a barely broken symmetry results in completely different behavior.

In summary, we found that adding a symmetry breaking term $-h_q\sum_x \cos(q\varphi_x)$ to the ordinary classical O(2) model results in a rich phase diagram containing second-order phase transitions of various universality classes and BKT transitions. This Extended-O(2) model can serve as a playground in which to explore these different kinds of phase transitions and their relationships. The model interpolates between different $\mathbb{Z}_q$ models via the continuously tunable parameter $q$, and so it may be useful for optimizing the various $\mathbb{Z}_q$ approximations of $U(1)$ in quantum simulations of Abelian gauge theories. Furthermore, since the O(2) model is a nontrivial limit of the Abelian-Higgs model, it may be useful to attempt the analog quantum simulation of the O(2) model and the extension of that model with the symmetry-breaking term as a first step toward the analog simulation of the Abelian-Higgs model.

\section*{Acknowledgements}

We thank Gerardo Ortiz, James Osborne, Nouman Butt, Richard Brower, Judah Unmuth-Yockey, and  members of the QuLAT collaboration for useful discussions and comments.

This work was supported in part by the U.S. Department of Energy (DOE) under Award Numbers DE-SC0010113, and DE-SC0019139. J.Z.\ is supported by
NSFC under Grants No.~12304172 and No.~12347101, Chongqing Natural Science Foundation under Grant No.~CSTB2023NSCQ-MSX0048, and Fundamental Research Funds for the Central Universities under Projects No.~2023CDJXY-048 and No. 2020CDJQY-Z003. This research used resources of the Syracuse University HTC Campus Grid and NSF award ACI-1341006 and the National Energy Research Scientific Computing Center (NERSC), a U.S. Department of Energy Office of Science User Facility located at Lawrence Berkeley National Laboratory, operated under Contract No. DE-AC02-05CH11231 using NERSC awards HEP-ERCAP0020659 and HEP-ERCAP0023235. This work was also supported in part through computational resources and services provided by the Institute for Cyber-Enabled Research at Michigan State University.

\bibliographystyle{apsrev4-1}


%

\appendix

\section{Connecting to the Clock Model}
\label{app:clockmodellimit}

The energy function of the ordinary $q$-state clock model with zero external field is
\begin{equation}
S = - \sum_{x,\mu}\cos(\varphi_{x+\hat\mu}-\varphi_{x}),
\end{equation}
where the allowed angles are the discrete values $\varphi_{x} = 2\pi k/q$ with $q,k\in \mathbb{Z}$. In a previous work \cite{hostetler2021}, we defined the ``Extended $q$-state clock model'' by allowing $q$ to take noninteger values and restricting the angles to
\begin{equation}
\varphi_0 \leq \varphi_{x} = \frac{2\pi k}{q} < \varphi_0 + 2\pi,
\end{equation}
with e.g. $\varphi_0 = 0$. In that work, we characterized the full phase diagram of the model.

In the present work, we consider the ``Extended-O(2) model'' with zero-field energy function
\begin{equation}
S = - \sum_{x,\mu}\cos(\varphi_{x+\hat\mu}-\varphi_{x}) - h_q\sum_x\cos(q\varphi_x) .
\end{equation}
Now, the allowed angles are the continuous values $\varphi\in[0,2\pi)$, but when $h_q$ is taken large, the Boltzmann factor $e^{-\beta S}$ forces the angles to take the values $\varphi_{x} = 2\pi k/q$ with $k\in \mathbb{Z}$. Hence, we claim that the Extended-O(2) model reduces to the Extended $q$-state clock model in the limit $h_q\rightarrow \infty$. However, there are some subtleties when $q$ is noninteger.

For noninteger $q$, the $2\pi$-rotational symmetry is broken. Thus, we have to carefully define the domain of the angles $\varphi_x$. For example, one could choose $\varphi_x \in [-\pi,\pi)$ or $\varphi_x \in [0,2\pi)$. These different choices lead to very different phase diagrams as we showed for the $h_q\rar\infty$ case in \cite{hostetler2021}. We prefer the choice $\varphi_x \in [0,2\pi)$, simply because the phase diagram gives a more consistent periodic picture as $q$ is increased\footnote{See the left panel of Figure 26 in \cite{hostetler2021} as opposed to the right panel in the same figure.}. However, with the choice $\varphi_x \in [0,2\pi)$, there is a hard cutoff at the angle $\varphi = 0$. When $h_q=\infty$, this is not a problem, but at finite $h_q$, this cutoff skews the distribution of angles severely enough that the model at finite $h_q$ does not smoothly connect to the model at infinite $h_q$ when $h_q\rar\infty$. To fix this, we need to slightly shift the angle domain such that $\varphi_x \in [-\varepsilon,2\pi-\varepsilon)$. To match the clock model (i.e. $h_q=\infty$ case), one needs an $\varepsilon$ that varies with $q$ and satisfies the condition $0<\varepsilon<2\pi(1-\lfloor q\rfloor/q)$. In this work, we choose
\begin{equation}
\varepsilon = \pi \left(1- \frac{\lfloor q\rfloor}{q} \right).
\end{equation}
After taking the $h_q\rar\infty$ limit, the $\varepsilon\rar 0$ limit can be taken to connect with the clock models.

When $q$ is noninteger, one may find that the Extended O(2) model with $h_q\rightarrow\infty$ and the Extended $q$-state clock model do not agree when $\beta\rightarrow 0$. For noninteger $q$, the $\mathbb{Z}_q$ symmetry is explicitly broken, and for example, the energy density of the Extended $q$-state clock model does not go to zero in the limit $\beta\rightarrow 0$ \cite{hostetler2021}. In contrast, the energy density in the Extended O(2) model does go to zero in the limit $\beta\rightarrow 0$ for all values of $h_q$. The discrepancy can be understood by considering the Boltzmann factor of the Extended O(2) model, which has the form
\begin{equation}
e^{-\beta S} = e^{\beta \sum\cos(\varphi_{x+\hat\mu}-\varphi_{x}) +\beta h_q\sum\cos(q\varphi_x)}
\end{equation}
The clock limit occurs when $h_q\rightarrow\infty$ and the second term becomes very large---forcing the angles into the discrete values $\varphi_{x} = 2\pi k/q$. When $\beta\rightarrow 0$, the two limits compete against each other. To match the clock model in the $h_q\rightarrow \infty$ limit for all values of $\beta$, one has to adjust $h_q$ such that $\beta h_q$ stays large as one approaches $\beta=0$.

\section{Finite-size Scaling Procedure}
\label{app:extractingmaxima}

\begin{figure}
\centering
\includegraphics[scale=.4]{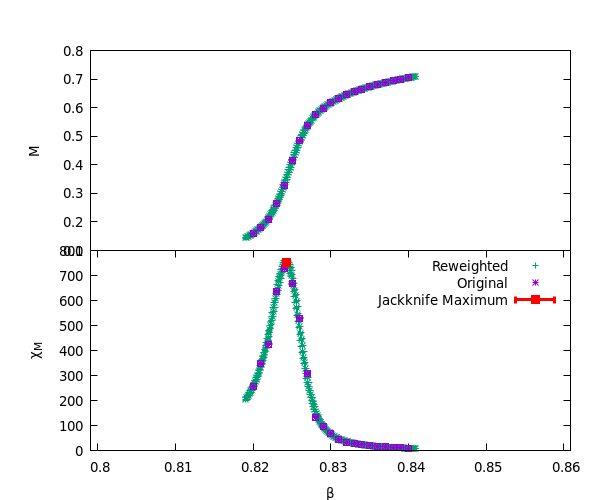}
\caption{An example with $q=3$, $h_q=1$, and lattice size $L=192$. The top panel shows the proxy magnetization, and the bottom panel shows the magnetic susceptibility. The purple points show the results from canonical simulations. The cubic fit to the peak is applied using the reweighted values shown in green. The red point with $x$ and $y$ errors shows the peak maximum as estimated by applying a cubic fit to each jackknife bin. \label{fullen_compare}}
\end{figure}

\begin{figure}
\centering
\includegraphics[scale=.4]{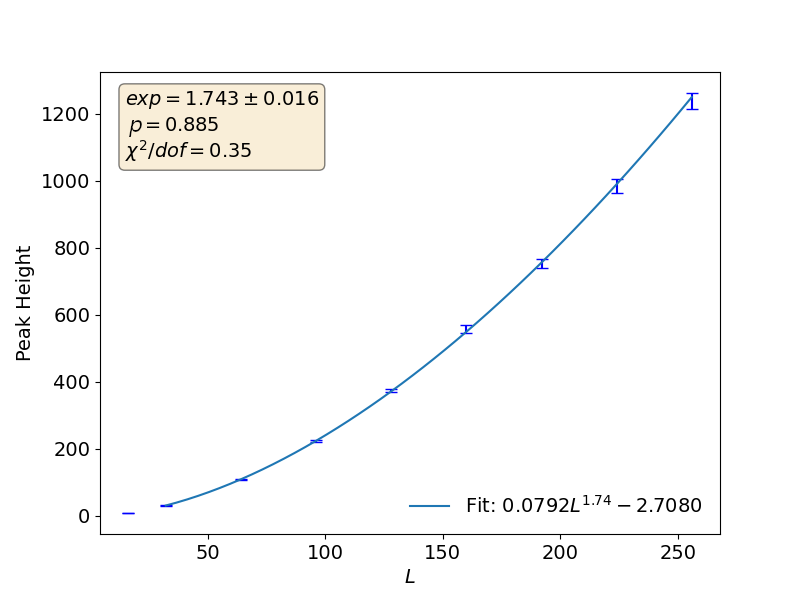}
\caption{An example of a finite size scaling fit for the model with $q=3$ and $h_q=1$. Here, we fit the maxima of the magnetic susceptibility to the ansatz given by Eq.~(\ref{eq:ansatz_magsus}).  The numbers shown here may not exactly match the values listed in Table~\ref{tab:criticalexponents} because a different averaging procedure is used there. \label{fullen_L}}
\end{figure}

\begin{figure}
\centering
\includegraphics[scale=.4]{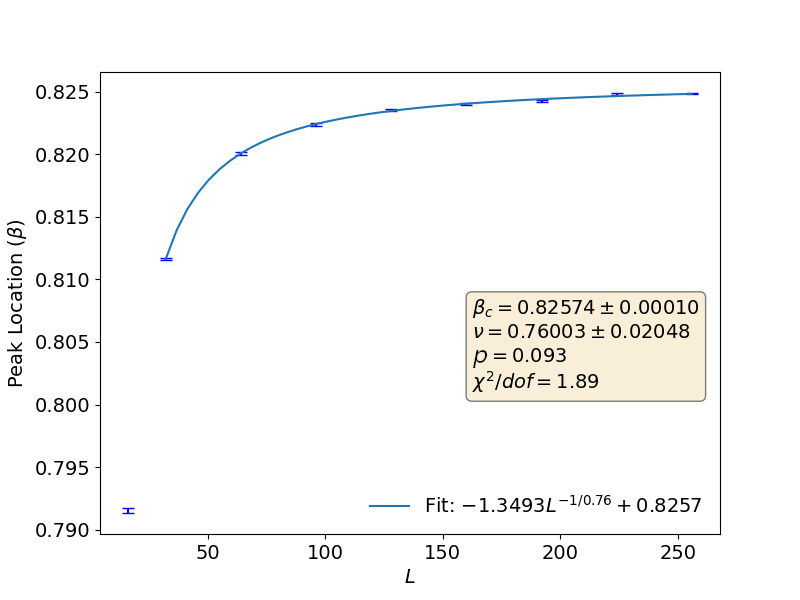}
\caption{An example of a finite size scaling fit for the model with $q=3$ and $h_q=1$. Here, we fit the $\beta$-location of the peaks of the magnetic susceptibility to the ansatz given by Eq.~(\ref{eq:ansatz_pc}). The numbers shown here may not exactly match the values listed in Table~\ref{tab:criticalpoints} because a different averaging procedure is used there. \label{txt_betacextrap}}
\end{figure}

We do canonical MCMC lattice simulations in the neighborhood of the critical point. We then use multihistogram reweighting to interpolate between these simulations to precisely identify the maximum of some quantity. For example, for the magnetic susceptibility, we:

\begin{enumerate}
\item Perform some number $N_{runs}$ of canonical MCMC lattice simulations. Each simulation gives us a time series of the proxy magnetization at a given $\beta$, and this is used to estimate the susceptibility for that $\beta$
\item Stitch these canonical runs together via multihistogram reweighting. See Fig.~\ref{fullen_compare}. This procedure gives us the magnetization estimated at many (i.e. $N_{\beta}>>N_{runs}$) $\beta$ values
	\begin{itemize}
	\item Before reweighting, the integrated autocorrelation time of the magnetization is estimated from each canonical time series
	\item The output of the reweighting procedure is $N_{jb}$ jackknife bins of the magnetization estimated at each of the $N_\beta$ $\beta$ values
	\end{itemize}
\item Within each jackknife bin, we compute the susceptibility at the $N_\beta$ $\beta$ values, fit a cubic polynomial to the peak of the susceptibility, and extract the maximum and the $\beta$-value at the maximum
\item In the end, we end up with $N_{jb}$ jackknife estimates of the maximum $\mathcal{M}$ of the susceptibility. See Fig.~\ref{fullen_compare}. Our best estimate of the maximum is
\begin{equation}
\overline{\mathcal{M}^J} \pm s_J (\mathcal{M}^J),
\end{equation}
where the jackknife mean and error bar are computed as
\begin{align}
\overline{\mathcal{M}^J} &= \frac{1}{N_{jb}} \sum_{i=1}^{N_{jb}} \mathcal{M}_i^J \\
s_J (\mathcal{M}^J) &= \sqrt{ \frac{N_{jb}-1}{N_{jb}}\sum_{i=1}^{N_{jb}} \left(\mathcal{M}_i^J - \overline{\mathcal{M}^J}\right)^2},
\end{align}
where $\mathcal{M}_i^J$ is the maximum computed from the $i$th jackknife bin. Note that ``$J$'' is merely a label to indicate that a quantity is from jackknife---it is not an index.
\item This procedure is repeated for several lattice sizes $L$ and then fit to the appropriate finite size scaling form Eq.~(\ref{eq:ansatz_deru})--(\ref{eq:ansatz_cv2}). See, for example, Figures~(\ref{fullen_L}) and (\ref{txt_betacextrap})
\end{enumerate}

\section{FSS Results at Integer \texorpdfstring{$q$}{q}}
\label{app:fss_results_intq}

\setlength{\tabcolsep}{0.5em} 
\renewcommand{\arraystretch}{1.1}
\begin{table*}
\centering
\begin{tabular}{ccccccccc} \hline
$q$ & $h_q$ & Exponent Ratios & Fit Form & $L_{min}$ & $L_{max}$ & $p$-value & $\chi^2/dof$ & Exponent Ratios\\ \hline
2.0 & 0.1 & $1/\nu$ & Eq.~(\ref{eq:ansatz_deru}) & 32 & 256 & 0.196 & 1.483 & 0.958(49) \\  
2.0 & 0.1 & $\alpha/\nu$ & Eq.~(\ref{eq:ansatz_cv2}) & 32 & 256 & 0.498 & 0.878 & 0.020(17) \\  
2.0 & 0.1 & $\gamma/\nu$ & Eq.~(\ref{eq:ansatz_magsus}) & 32 & 256 & 0.386 & 1.058 & 1.780(16) \\  
2.0 & 0.1 & $-\beta/\nu$ & Eq.~(\ref{eq:ansatz_inflM}) & 32 & 256 & 0.617 & 0.711 & -0.306(81) \\  
2.0 & 0.1 & $2-\eta$ & Eq.~(\ref{eq:ansatz_sf}) & 32 & 256 & 0.198 & 1.476 & 1.713(13) \\  
2.0 & 1.0 & $1/\nu$ & Eq.~(\ref{eq:ansatz_deru}) & 32 & 256 & 0.722 & 0.571 & 0.978(68) \\  
2.0 & 1.0 & $\alpha/\nu$ & Eq.~(\ref{eq:ansatz_cv2}) & 32 & 256 & 0.011 & 2.991 & -0.010(16) \\  
2.0 & 1.0 & $\gamma/\nu$ & Eq.~(\ref{eq:ansatz_magsus}) & 32 & 256 & 0.532 & 0.828 & 1.745(19) \\  
2.0 & 1.0 & $-\beta/\nu$ & Eq.~(\ref{eq:ansatz_inflM}) &  &  &  &  &  \\  
2.0 & 1.0 & $2-\eta$ & Eq.~(\ref{eq:ansatz_sf}) & 32 & 256 & 0.07 & 2.06 & 1.749(14) \\ 
3.0 & 0.1 & $1/\nu$ & Eq.~(\ref{eq:ansatz_deru}) & 32 & 256 & 0.858 & 0.384 & 1.521(57) \\  
3.0 & 0.1 & $\alpha/\nu$ & Eq.~(\ref{eq:ansatz_cv}) & 32 & 256 & 0.029 & 2.514 & 0.538(48) \\  
3.0 & 0.1 & $\gamma/\nu$ & Eq.~(\ref{eq:ansatz_magsus}) & 32 & 256 & 0.264 & 1.302 & 1.962(16) \\ 
3.0 & 0.1 & $-\beta/\nu$ & Eq.~(\ref{eq:ansatz_inflM}) &  &  &  &  &  \\  
3.0 & 0.1 & $2-\eta$ & Eq.~(\ref{eq:ansatz_sf}) & 32 & 256 & 0.852 & 0.392 & 1.6349(93) \\  
3.0 & 1.0 & $1/\nu$ & Eq.~(\ref{eq:ansatz_deru}) & 32 & 256 & 0.822 & 0.436 & 1.236(40) \\  
3.0 & 1.0 & $\alpha/\nu$ & Eq.~(\ref{eq:ansatz_cv}) & 32 & 256 & 0.402 & 1.032 & 0.385(21) \\  
3.0 & 1.0 & $\gamma/\nu$ & Eq.~(\ref{eq:ansatz_magsus}) & 32 & 256 & 0.873 & 0.361 & 1.743(16) \\  
3.0 & 1.0 & $-\beta/\nu$ & Eq.~(\ref{eq:ansatz_inflM}) &  &  &  &  &  \\  
3.0 & 1.0 & $2-\eta$ & Eq.~(\ref{eq:ansatz_sf}) & 32 & 256 & 0.407 & 1.021 & 1.705(16) \\ 
4.0 & 0.1 & $1/\nu$ & Eq.~(\ref{eq:ansatz_deru}) & 32 & 256 & 0.942 & 0.241 & 0.40(14) \\  
4.0 & 0.1 & $\alpha/\nu$ & Eq.~(\ref{eq:ansatz_cv2}) & 32 & 256 & 0 & 10.332 & -0.120(19) \\  
4.0 & 0.1 & $\gamma/\nu$ & Eq.~(\ref{eq:ansatz_magsus}) & 32 & 256 & 0.53 & 0.831 & 1.728(16) \\  
4.0 & 0.1 & $-\beta/\nu$ & Eq.~(\ref{eq:ansatz_inflM}) & 32 & 256 & 0.09 & 1.922 & -0.31(31) \\  
4.0 & 0.1 & $2-\eta$ & Eq.~(\ref{eq:ansatz_sf}) & 32 & 256 & 0.398 & 1.035 & 1.7430(85) \\  
4.0 & 1.0 & $1/\nu$ & Eq.~(\ref{eq:ansatz_deru}) & 32 & 256 & 0.167 & 1.576 & 0.834(96) \\  
4.0 & 1.0 & $\alpha/\nu$ & Eq.~(\ref{eq:ansatz_cv2}) & 32 & 256 & 0.53 & 0.83 & -0.1349(90) \\  
4.0 & 1.0 & $\gamma/\nu$ & Eq.~(\ref{eq:ansatz_magsus}) & 32 & 256 & 0.568 & 0.777 & 1.742(14) \\  
4.0 & 1.0 & $-\beta/\nu$ & Eq.~(\ref{eq:ansatz_inflM}) & 32 & 256 & 0.39 & 1.049 & -0.65(51) \\  
4.0 & 1.0 & $2-\eta$ & Eq.~(\ref{eq:ansatz_sf}) & 32 & 256 & 0.073 & 2.032 & 1.7468(93) \\ 
5.0 & 0.1 & $\gamma/\nu$ & Eq.~(\ref{eq:ansatz_magsus}) & 32 & 256 & 0.039 & 2.379 & 1.727(15) \\  
5.0 & 0.1 & $-\beta/\nu$ & Eq.~(\ref{eq:ansatz_inflM}) & 32 & 256 & 0.067 & 2.083 & -0.35(22) \\  
5.0 & 0.1 & $2-\eta$ & Eq.~(\ref{eq:ansatz_sf}) & 32 & 256 & 0.028 & 2.541 & 1.7284(72) \\  
5.0 & 1.0 & $\gamma/\nu$ & Eq.~(\ref{eq:ansatz_magsus}) & 32 & 256 & 0.042 & 2.339 & 1.722(14) \\  
5.0 & 1.0 & $-\beta/\nu$ & Eq.~(\ref{eq:ansatz_inflM}) & 32 & 256 & 0.06 & 2.148 & -0.44(24) \\  
5.0 & 1.0 & $2-\eta$ & Eq.~(\ref{eq:ansatz_sf}) & 32 & 256 & 0.354 & 1.117 & 1.7347(94) \\ 
6.0 & 0.1 & $\gamma/\nu$ & Eq.~(\ref{eq:ansatz_magsus}) & 32 & 256 & 0.678 & 0.629 & 1.710(11) \\  
6.0 & 0.1 & $-\beta/\nu$ & Eq.~(\ref{eq:ansatz_inflM}) & 32 & 256 & 0.009 & 3.08 & -0.15(16) \\  
6.0 & 0.1 & $2-\eta$ & Eq.~(\ref{eq:ansatz_sf}) & 32 & 256 & 0.214 & 1.432 & 1.7120(86) \\  
6.0 & 1.0 & $\gamma/\nu$ & Eq.~(\ref{eq:ansatz_magsus}) & 32 & 256 & 0.867 & 0.371 & 1.726(15) \\  
6.0 & 1.0 & $-\beta/\nu$ & Eq.~(\ref{eq:ansatz_inflM}) &  &  &  &  &  \\  
6.0 & 1.0 & $2-\eta$ & Eq.~(\ref{eq:ansatz_sf}) & 32 & 256 & 0.096 & 1.889 & 1.7346(92) \\ \hline 
\end{tabular}
\caption{Critical exponents are tabulated for integer $q=2,3,4,5,6$ and finite $h_q=0.1,1.0$. The $p$-value is a goodness-of-fit measure defined in Eq.~(\ref{fitq}). For $q=5,6$, there appear to be two BKT transitions as is the case for $h_q=\infty$. Here, we include data only for the transition at small inverse temperature. Empty cells indicate that it was not possible to obtain a trustworthy fit.}
\label{tab:criticalexponents}
\end{table*}

\setlength{\tabcolsep}{0.5em} 
\renewcommand{\arraystretch}{1.1}
\begin{table*}
\centering
\begin{tabular}{ccccccccc} \hline
$q$ & $h_q$ & From & $L_{min}$ & $L_{max}$ & $p$-value & $\chi^2/dof$ & $\nu$ & $\beta_c$ \\ \hline 
2.0 & 0.1 & $\left.{dU_{M}}/{d\beta}\right|_{max}$ & 32 & 256 & 0.119 & 1.767 & 1.13(10) & 0.8775(12) \\  
2.0 & 0.1 & $\left.C_V\right|_{max}$ & 32 & 256 & 0.016 & 2.816 & 1.03(11) & 0.87726(38) \\  
2.0 & 0.1 & $\left.\chi_{M}\right|_{max}$ & 32 & 256 & 0.256 & 1.32 & 1.054(54) & 0.87753(48) \\  
2.0 & 0.1 & $\left.\langle M\rangle\right|_{infl}$ & 32 & 256 & 0.172 & 1.561 & 1.014(80) & 0.87734(49) \\  
2.0 & 0.1 & $\left.F(\vec{q})\right|_{max}$ & 32 & 256 & 0.179 & 1.534 & 1.066(45) & 0.87816(76) \\  
2.0 & 1.0 & $\left.{dU_{M}}/{d\beta}\right|_{max}$ & 32 & 256 & 0.123 & 1.752 & 0.979(73) & 0.65440(53) \\  
2.0 & 1.0 & $\left.C_V\right|_{max}$ & 32 & 256 & 0.164 & 1.59 & 1.04(16) & 0.65442(27) \\  
2.0 & 1.0 & $\left.\chi_{M}\right|_{max}$ & 32 & 256 & 0.842 & 0.407 & 0.991(55) & 0.65453(28) \\  
2.0 & 1.0 & $\left.\langle M\rangle\right|_{infl}$ & 32 & 256 & 0.963 & 0.194 & 0.908(82) & 0.65423(28) \\  
2.0 & 1.0 & $\left.F(\vec{q})\right|_{max}$ & 32 & 256 & 0.778 & 0.497 & 1.012(42) & 0.65482(40) \\ 
3.0 & 0.1 & $\left.{dU_{M}}/{d\beta}\right|_{max}$ & 32 & 256 & 0.133 & 1.706 & 0.791(35) & 0.99873(93) \\  
3.0 & 0.1 & $\left.C_V\right|_{max}$ & 32 & 256 & 0.924 & 0.276 & 0.863(41) & 1.00073(74) \\  
3.0 & 0.1 & $\left.\chi_{M}\right|_{max}$ & 32 & 256 & 0.03 & 2.507 & 0.811(19) & 0.99924(45) \\  
3.0 & 0.1 & $\left.\langle M\rangle\right|_{infl}$ & 32 & 256 & 0.176 & 1.545 & 0.756(29) & 0.99878(54) \\  
3.0 & 0.1 & $\left.F(\vec{q})\right|_{max}$ & 32 & 256 & 0.215 & 1.429 & 1.168(44) & 1.0047(16) \\  
3.0 & 1.0 & $\left.{dU_{M}}/{d\beta}\right|_{max}$ & 32 & 256 & 0.22 & 1.415 & 0.808(31) & 0.82568(18) \\  
3.0 & 1.0 & $\left.C_V\right|_{max}$ & 32 & 256 & 0.173 & 1.558 & 0.909(47) & 0.82603(13) \\  
3.0 & 1.0 & $\left.\chi_{M}\right|_{max}$ & 32 & 256 & 0.093 & 1.905 & 0.760(21) & 0.82574(10) \\  
3.0 & 1.0 & $\left.\langle M\rangle\right|_{infl}$ & 32 & 256 & 0.12 & 1.764 & 0.760(27) & 0.82575(11) \\  
3.0 & 1.0 & $\left.F(\vec{q})\right|_{max}$ & 32 & 256 & 0.015 & 2.855 & 0.823(18) & 0.82603(19) \\
4.0 & 0.1 & $\left.{dU_{M}}/{d\beta}\right|_{max}$ & 32 & 256 & 0.158 & 1.607 & 2.26(59) & 1.098(27) \\  
4.0 & 0.1 & $\left.C_V\right|_{max}$ &  &  &  &  &  &  \\  
4.0 & 0.1 & $\left.\chi_{M}\right|_{max}$ & 32 & 256 & 0.229 & 1.387 & 1.96(12) & 1.0797(54) \\  
4.0 & 0.1 & $\left.\langle M\rangle\right|_{infl}$ & 32 & 256 & 0.082 & 1.972 & 2.05(26) & 1.082(12) \\  
4.0 & 0.1 & $\left.F(\vec{q})\right|_{max}$ & 32 & 256 & 0.839 & 0.412 & 1.88(13) & 1.0772(67) \\  
4.0 & 1.0 & $\left.{dU_{M}}/{d\beta}\right|_{max}$ & 32 & 256 & 0.954 & 0.216 & 1.34(17) & 0.9918(47) \\  
4.0 & 1.0 & $\left.C_V\right|_{max}$ &  &  &  &  &  &  \\  
4.0 & 1.0 & $\left.\chi_{M}\right|_{max}$ & 32 & 256 & 0.053 & 2.21 & 1.251(47) & 0.9883(13) \\  
4.0 & 1.0 & $\left.\langle M\rangle\right|_{infl}$ & 32 & 256 & 0.46 & 0.935 & 1.52(26) & 0.9952(60) \\  
4.0 & 1.0 & $\left.F(\vec{q})\right|_{max}$ & 32 & 256 & 0.791 & 0.479 & 1.334(67) & 0.9912(20) \\ \hline 
\end{tabular}
\caption{Here we fit the peak (or inflection point) locations to the form Eq.~(\ref{eq:ansatz_pc}) to estimate the infinite-volume critical point $\beta_c$ and the critical exponent $\nu$. The $p$-value is a goodness-of-fit measure defined in Eq.~(\ref{fitq}). Empty cells indicate that it was not possible to obtain a trustworthy fit.}
\label{tab:criticalpoints}
\end{table*}

\setlength{\tabcolsep}{0.5em} 
\renewcommand{\arraystretch}{1.1}
\begin{table*}
\centering
\begin{tabular}{cccccccc} \hline
$q$ & $h_q$ & From & $L_{min}$ & $L_{max}$ & $p$-value & $\chi^2/dof$ & $\beta_c$ \\ \hline
5.0 & 0.1 & $\left.\chi_{M}\right|_{max}$ & 32 & 256 & 0.167 & 1.579 & 1.1308(77) \\  
5.0 & 0.1 & $\left.\langle M\rangle\right|_{infl}$ & 32 & 256 & 0.037 & 2.402 & 1.143(23) \\  
5.0 & 0.1 & $\left.F(\vec{q})\right|_{max}$ & 32 & 256 & 0.26 & 1.312 & 1.133(13) \\  
5.0 & 1.0 & $\left.\chi_{M}\right|_{max}$ & 32 & 256 & 0.571 & 0.774 & 1.1267(96) \\  
5.0 & 1.0 & $\left.\langle M\rangle\right|_{infl}$ & 32 & 256 & 0.078 & 2.007 & 1.133(22) \\  
5.0 & 1.0 & $\left.F(\vec{q})\right|_{max}$ & 32 & 256 & 0.939 & 0.246 & 1.126(15) \\ 
6.0 & 0.1 & $\left.\chi_{M}\right|_{max}$ & 32 & 256 & 0.11 & 1.819 & 1.1237(76) \\  
6.0 & 0.1 & $\left.\langle M\rangle\right|_{infl}$ & 32 & 256 & 0.012 & 2.971 & 1.123(13) \\  
6.0 & 0.1 & $\left.F(\vec{q})\right|_{max}$ & 32 & 256 & 0.348 & 1.127 & 1.128(12) \\  
6.0 & 1.0 & $\left.\chi_{M}\right|_{max}$ & 32 & 256 & 0.209 & 1.445 & 1.125(11) \\  
6.0 & 1.0 & $\left.\langle M\rangle\right|_{infl}$ & 32 & 256 & 0.339 & 1.143 & 1.099(16) \\  
6.0 & 1.0 & $\left.F(\vec{q})\right|_{max}$ & 32 & 256 & 0.206 & 1.456 & 1.118(13) \\ \hline 
\end{tabular}
\caption{Here we fit the peak (or inflection point) locations to the form Eq.~(\ref{eq:ansatz_pc_bkt2}) to estimate the infinite-volume critical point $\beta_c$ of the small-$\beta$ phase transition. The $p$-value is a goodness-of-fit measure defined in Eq.~(\ref{fitq}).}
\label{tab:criticalpointsbkt}
\end{table*}

\clearpage
\section{TRG and MC Comparisons}
\label{app:mcvstrg}

Whereas Monte Carlo is a well-established approach with easily quantifiable uncertainties, it is an approach that sometimes struggles with critical slowing down. In the Extended-O(2) model studied in this paper, the Monte Carlo approach has no problem when $q$ is integer. However, when $q$ is noninteger, the Markov chain suffers from large autocorrelation times which worsen with increasing lattice size. The result is that for noninteger $q$, it becomes infeasible to produce de-correlated samples for large volumes e.g. for $L>>32$.

In contrast, tensor renormalization group (TRG) methods do not suffer from autocorrelation and critical slowing down. However, because of the truncations used, there are systematic uncertainties that are not always well-quantified. To be confident that our TRG methods are reliable, we compare them against MC in the regimes which are accessible to Monte Carlo.

In Fig.~\ref{fig:mcvstrg_comparison}, we consider several different values of $q$ and $h_q$ at two different volumes. For noninteger values of $q$, we compare small volumes $L=16,32$. For integer values of $q$, we are able to compare larger volumes $L=128,256$. In the top panel of each figure, we compare the specific heat from Monte Carlo and TRG. With TRG, we compute the free energy and apply smoothing splines before computing the specific heat as the second derivative of the free energy. In the bottom panel, we show the magnetic susceptibility from Monte Carlo. These figures show very good agreement between Monte Carlo and TRG for the specific heat.

With Monte Carlo, we use zero external field ($h=0$), and so the magnetization $\vec{M}$ averaged over many equilibrium configurations would average to zero for all $\beta$. Instead, we measure a proxy magnetization $|\vec{M}|$ defined in Eq.~(\ref{eq_proxymag}). The corresponding susceptibility is defined in Eq.~(\ref{eq_proxyc}).

With TRG, we use a different approach. 
Since the TRG procedure preserves the symmetry, zero external field yields zero magnetization.
Thus in this study we imposed the finite external field to extract the magnetization.
In a technical point of view we have used the method proposed by Morita and Kawashima in ref.~\cite{Morita2019} to calculate the magnetization (as declared in the method section).
Note that the magnetizations and the susceptibilities calculated by the TRG are based on the normal magnetization rather than the proxy one. 
However, again, we expect the critical behavior does not change between the normal and proxy definitions.

\begin{figure}
\centering
\vspace{0.1cm} 
\includegraphics[width=\columnwidth]{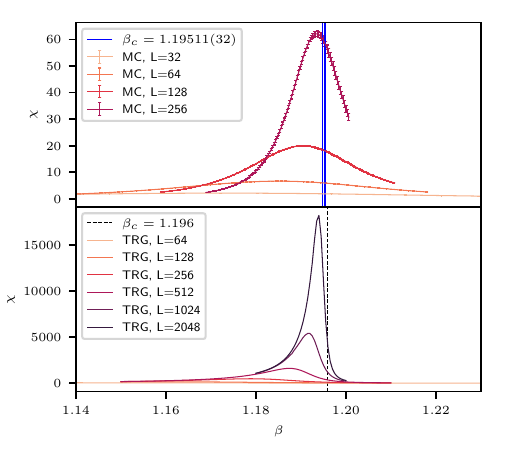}
\caption{We compute the susceptibility for $q=3.9$ and $h_q=1$ in the vicinity of the second-order transition in two different ways. In the top panel, we see how the susceptibility peaks from Monte Carlo approach the critical point, and in the bottom panel we look at the susceptibility peaks from TRG. From the Monte Carlo peak positions, we estimate $\beta_c = 1.19511(32)$ (vertical blue line) by fitting the peak locations to the finite-size-scaling form Eq.~(\ref{eq:ansatz_pc}). From the TRG peak positions, we estimate $\beta_c=1.196$ (vertical dashed line) via the same finite-size scaling form. As described in the main text, because of large autocorrelations, the Monte Carlo results at noninteger $q$ are not strictly reliable for $L>32$, and this might explain the small observed discrepancy between MC and TRG. \label{mcvstrg_q39g1_zoomed_unrescaled}}
\end{figure}

\begin{figure*}
        \centering
        \begin{subfigure}[b]{0.42\textwidth}
            \centering
            \includegraphics[width=\textwidth]{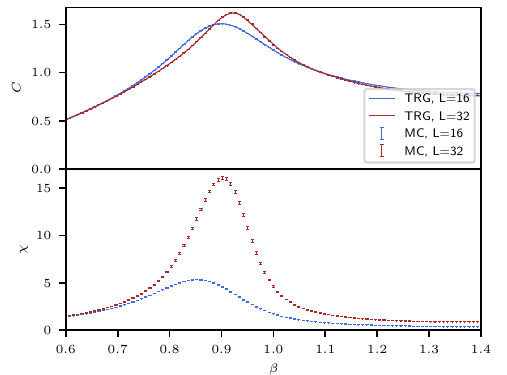}
            \caption{{ $q=3.9$, $h_q=0.1$}}
        \end{subfigure}
        \hfill
        \begin{subfigure}[b]{0.42\textwidth}  
            \centering 
            \includegraphics[width=\textwidth]{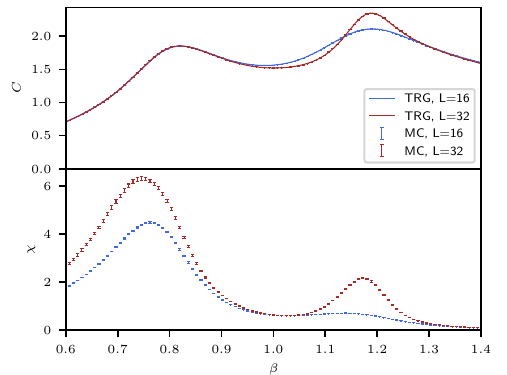}
            \caption{{ $q=3.9$, $h_q=1$}}   
        \end{subfigure}
        \begin{subfigure}[b]{0.42\textwidth}   
            \centering 
            \includegraphics[width=\textwidth]{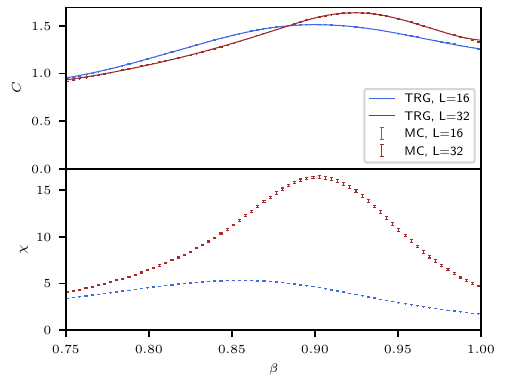}
            \caption{{ $q=4.1$, $h_q=0.1$}}   
        \end{subfigure}
        \hfill
        \begin{subfigure}[b]{0.42\textwidth}   
            \centering 
            \includegraphics[width=\textwidth]{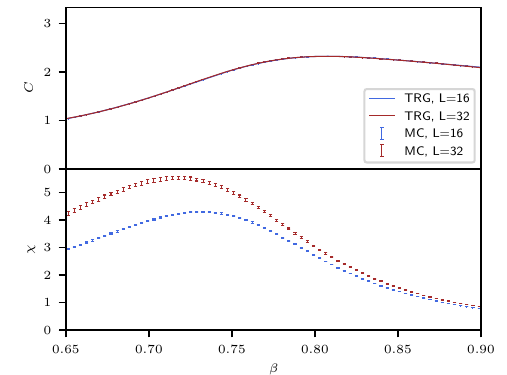}
            \caption{{ $q=4.1$, $h_q=1$}}    
        \end{subfigure}
        \begin{subfigure}[b]{0.42\textwidth}   
            \centering 
            \includegraphics[width=\textwidth]{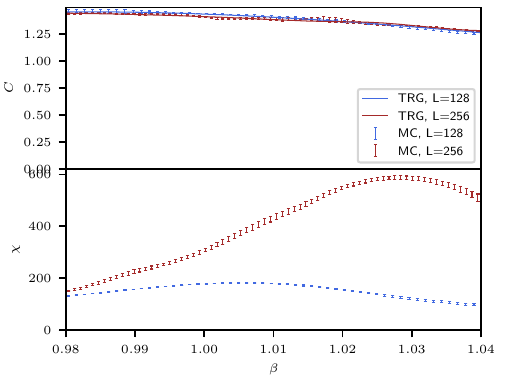}
            \caption{{ $q=5$, $h_q=0.1$}}   
        \end{subfigure}
        \hfill
        \begin{subfigure}[b]{0.42\textwidth}   
            \centering 
            \includegraphics[width=\textwidth]{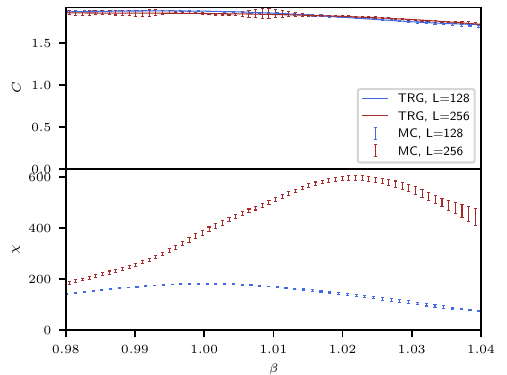}
            \caption{{ $q=5$, $h_q=1$}}    
        \end{subfigure}
        \caption{{We compare Monte Carlo and TRG results at various model parameters. In each figure, we compare in the top panel the specific heat $C$ from Monte Carlo (error bars and no connecting lines) with the specific heat from TRG (solid line). In the bottom panel of each figure, we show the magnetic susceptibility $\chi$ from Monte Carlo. The TRG susceptibility is not shown since the susceptibility cannot be directly compared for the two methods. For the top two figures, we show the case $q=3.9$ with $h_q=0.1$ (left) and $h_q=1$ (right). The $\beta$ range was chosen to show both peaks in the specific heat, although at these small volumes, only one peak is visible for $h_q=0.1$. In the middle two figures, we show the case $q=4.1$ with $h_q=0.1$ (left) and $h_q=1$ (right). The $\beta$ range was chosen to show small-$\beta$ peak in the specific heat. For the case $h_q=1$, the specific heats of the two different volumes lie directly on top of each other. In the bottom two figures, we show the case $q=5$ with $h_q=0.1$ (left) and $h_q=1$ (right). At integer $q$, we are able to reliably go to larger volumes with Monte Carlo, so we compare $L=128$ and $L=256$ here. The $\beta$ range was chosen to show the small-$\beta$ peak in the specific heat.}}
        \label{fig:mcvstrg_comparison}
    \end{figure*}

Because of the different approaches used by Monte Carlo and TRG, it is not meaningful to directly compare the magnetization or the magnetic susceptibility. Hence, we do not include TRG susceptibilities in the bottom panels in Fig.~\ref{fig:mcvstrg_comparison}. However, we expect the critical behavior to be the same in the thermodynamic limit. We show an example of this in Fig.~\ref{mcvstrg_q39g1_zoomed_unrescaled} for the case $q=3.9$ and $\gamma=1$. In the top panel, we show how the susceptibility (calculated from Monte Carlo) peaks approach the critical point with increasing volume. The critical point $\beta_c = 1.19511(32)$ (vertical blue line) was estimated by fitting the Monte Carlo peak locations to the finite-size scaling form Eq.~(\ref{eq:ansatz_pc}). In the bottom panel of Fig.~\ref{mcvstrg_q39g1_zoomed_unrescaled}, we do the same but with the susceptibility calculated using TRG. Here, we estimate $\beta_c=1.196$ (vertical dashed line) using the same finite-size scaling form. Note, that the Monte Carlo results are not strictly reliable for $L>32$ in this figure since for $q=3.9$ there are very large and unmitigated autocorrelations. Nevertheless, it seems the finite-size scaling shift of the MC peaks are in general agreement with the finite-size scaling shift of the TRG peaks.

\end{document}